\documentclass[iop]{emulateapj}
\usepackage{graphicx}
\usepackage{multirow}
\newcommand{\nar}{NewAR}
\newcommand{\actaa}{AcA}

\slugcomment{Accepted to the Astrophysical Journal 4/2013}

\shorttitle{Identification of 5 GBS Interacting Binaries}
\shortauthors{Britt et al.}

\begin{document}
\title{Identification of 5 Interacting Binaries in the Galactic Bulge Survey}

\author{C.T. Britt\altaffilmark{1,2}, M.A.P. Torres \altaffilmark{3,4}, R.I. Hynes\altaffilmark{1,2}, P.G. Jonker\altaffilmark{3,4,5}, T.J. Maccarone\altaffilmark{6}, S. Greiss\altaffilmark{7}, D. Steeghs\altaffilmark{7,4}, P. Groot\altaffilmark{8,5}, C. Knigge \altaffilmark{6}, A. Dieball\altaffilmark{6}, G. Nelemans \altaffilmark{5}, V.J. Mikles \altaffilmark{1}, L. Gossen  \altaffilmark{1,2}}

\altaffiltext{1}{Louisiana State University, Department of Physics and Astronomy, Baton Rouge, LA 70803-4001, USA} \altaffiltext{2}{Visiting astronomer, Cerro Tololo Inter-American Observatory, National Optical Astronomy Observatory, which are operated by the Association of Universities for Research in Astronomy, under contract with the National Science Foundation.} \altaffiltext{3}{SRON, Netherlands Institute for Space Research, Sorbonnelaan 2, 3584 CA, Utrecht, The Netherlands} \altaffiltext{4}{Harvard-Smithsonian Center for Astrophysics, 60 Garden Street, Cambridge, MA 02138, USA} \altaffiltext{5}{Dept. of Astrophysics/IMAPP, Radboud University Nijmegen, P.O. Box 9010, 6500 GL, Nijmegen, The Netherlands}\altaffiltext{6}{School of Physics and Astronomy, University of Southampton, Hampshire SO17 1BJ, United Kingdom}\altaffiltext{7}{Astronomy and Astrophysics, Dept. of Physics, University of Warwick, Coventry, CV4 7AL, United Kingdom}\altaffiltext{8}{Cahill Center for Astronomy and Astrophysics, California Institute of Technology, 1200 East California Boulevard, Pasadena, CA 91125, USA}

\begin{abstract}
We present optical lightcurves, spectroscopy, and classification of five X-ray sources in the Chandra Galactic Bulge Survey [CXOGBS J174009.1-284725 (CX5), CXOGBS J173935.7-272935 (CX18), CXOGBS J173946.9-271809 (CX28), CXOGBS J173729.1-292804 (CX37), CXOGBS J174607.6-261547 (CX561)]. These objects were selected based on bright optical counterparts which were quickly found to have emission lines in their optical spectra. This paper presents an illustration of GBS optical follow-up, targeting emission line objects. Of these five objects, four exhibit photometric variability in the Sloan $r'$ band. CX5 shows a tentative period of 2.1 hours and is clearly an Intermediate Polar (IP).  CX28 and CX37 both exhibit flickering with no clear period. Both are suggested to also be IPs. CX18 was observed to undergo 2 dwarf nova outbursts. Finally, CX561 shows no detectable variability, although its characteristics would be consistent with either a quiescent Low Mass X-ray Binary or Cataclysmic Variable. 

\end{abstract}

\keywords{Surveys --- X-rays: binaries --- stars : emission-line --- binaries: close --- Stars: dwarf novae --- Novae, cataclysmic variables}

\section{Introduction}

The most comprehensive Galactic X-ray surveys of faint sources have focused on the Galactic Center or in globular clusters. Surveys of the Galactic Center carry the advantage of high source density \citep{Muno03}, but also the disadvantages for optical follow up of high crowding and $A_V$ on the order of 30 magnitudes. Together, those disadvantages make the determination of optical or infrared counterparts to X-ray sources in the Galactic Center very difficult \citep{Mauerhan09}. Establishing an optical/infrared counterpart is a necessary first step for characterizing the properties of the X-ray emitting objects. This characterization is done through a combination of the ratio of X-ray to optical luminosities, detection of ellipsoidal modulations of the companion, and optical and infrared spectroscopy. Studies of globular clusters escape the problem of high extinction, but the crowding problem is even more severe. Also, X-ray Binary (XRB) formation in globular clusters is dominated by dynamical processes, so they do not provide a probe of binary evolution in the field. 

The Galactic Bulge Survey (GBS) is intended to avoid as much as possible the problems of crowding and extinction present in previous surveys of the Galactic Center, while giving up as little as possible in the way of number of sources \citep{jon11}. The GBS makes use of both optical and X-ray imaging of two $6^{\circ} \times 1^{\circ}$ strips located $1.5^{\circ}$ above and below the Galactic plane, cutting out the region with $|b|<1^{\circ}$ to avoid copious amounts of dust in the Galactic Plane. Other surveys, such as the ChaMPlane survey of bright X-ray sources, are also having some success in identifying sources along the plane and in low extinction windows \citep{champlane12,vandenberg09}. The GBS is a shallower survey than others, with exposures of ~2ks \citep{jon11}. This was done because deeper observations would pick up disproportionately more Cataclysmic Variables (CVs) than X-ray Binaries (XRBs) and also would overshoot the ability to do optical spectroscopy on candidate counterparts \citep{jon11}.

There are multiple goals to be achieved in such a census of X-ray sources \citep{jon11}. We aim to greatly expand the known number of Galactic XRBs. By increasing the number of known low mass X-ray binaries (LMXBs), we are bound to correspondingly increase the number of LMXBs for which mass determinations are possible.  
Identification of source class types is another main goal of the GBS because it allows constraints to be placed on binary evolution models by comparing observed source class numbers to the predictions of population synthesis models. Such models are widely divergent in their predictions \citep{Ivanova05,Kalogera99,Pfahl03}. Underlying these predictions are assumptions concerning the common envelope phase of binary evolution \citep{Taam00}. This paper presents the first few sources to be identified in the GBS as new interacting binaries.

We expect to find a large number of CVs in the survey, both with and without strong magnetic fields \citep{jon11}. In the absence of a strong magnetic field, the white dwarf (WD) rotation period is not tidally locked to the binary orbital period (unlike the rotation period of the companion), the magnetic pressure never dominates the accretion flow, and an accretion disk can form. While these systems produce X-rays, the boundary layer in high accretion rate systems has a high optical depth that quenches X-ray emission, reprocessing it into UV light \citep{Warner03}. Non-magnetic CVs detected by the GBS are most likely quiescent CVs, where the amount of energy released as X-ray radiation is roughly two orders of magnitude less than the optical light from the system ($\frac{L_{X}}{L_{Opt}} \sim \frac{F_X}{F_{Opt}} \sim \frac{1}{100}$), though it can be higher in systems with higher accretion rates (up to $\dot{M} \approx 10^{16} {\rm g\,s^{-1}}$). In general, $\frac{F_X}{F_{Opt}}$ is correlated with both HeII 4686 and H$\beta$ emission, but more strongly with HeII 4686 \citep{Grindlay99}. 

A stronger magnetic field can lift material in accretion curtains off of the disk and funnel the material onto the magnetic pole. If the field is not strong enough to lock the WD's spin to the orbital period of the system, it is called a DQ Her system or an Intermediate Polar (IP).  Magnetic WD systems can produce a great deal more X-ray light than ordinary CVs, with $\frac{F_X}{F_{Opt}} \sim 1$ \citep{Warner03,Patterson94}. In IPs, since the magnetic pole of the WD is likely not aligned with the spin axis, the bright accretion spot forms a beam \citep{Patterson94}. X-ray light from IPs is therefore pulsed on the rotation period of the WD. Some of this beam will hit structures orbiting the WD, which reprocess the X-ray light. In the optical, therefore, the period for reprocessed light is often observed.  This is called the ``orbital sideband'' period, and is given by $\omega_{\mathrm{Side Band}} = \omega_{\mathrm{Spin}}-\Omega_{\mathrm{Orb}}$. 

\begin{table*}
\begin{center}
\caption{}
\begin{tabular}{ c c c c c l c}
\hline
\hline
Category & $\frac{F_{x}}{F_{Opt}}$ & $\frac{{\rm EW(HeI 4471)}}{{\rm EW(H\beta)}}$ & $\frac{{\rm EW(HeI 6678)}}{{\rm EW(H\beta)}}$ & $\frac{{\rm EW(HeII 4686)}}{{\rm EW(H\beta)}}$  & Possible Variability & References\\
\hline
CV & $0.01 \,\mbox{-}\, 1$ & $0.22 \pm 0.09$\footnotemark[1] & 0.1-0.5 & $<0.4$\footnotemark[2] & Flickering, sinusoidal, ellipsoidal, DN  & 1,2,3,11,13\\
IP & $0.1 \,\mbox{-}\, 10$ & $0.17 \pm 0.04$\footnotemark[1] & 0.1-0.5 & $>0.4$\footnotemark[2] & Flickering plus reprocessed X-ray pulsation,& 1,2,3,11,12 \\
 & & & & & sometimes DN & \\
qLMXB (NS) & $0.1\,\mbox{-}\, 1$ & nondetected \footnotemark[3] & $0.12$\footnotemark[3] & nondetected \footnotemark[3] & Flickering, ellipsoidal, flares on timescale of & 7,8 \\ 
 & & & & & tens of minutes or longer & \\
qLMXB (BH) & $0.01 \,\mbox{-}\, 0.1$ & nondetected \footnotemark[3] & $0.12$\footnotemark[3] & nondetected \footnotemark[3] & Flickering, ellipsoidal, flares on timescale of& 5,7,8\\
 & & & & &  tens of minutes or longer & \\
LMXB & $\ge 100$ & $\le 0.1$ \footnotemark[3] & 0.3\footnotemark[3] & 0.8\footnotemark[3] & Flickering, disk dominated, outbursts on & 4,6,9,10\\
 & & & & & timescale of a week to months, reprocessed  & \\
 & & & & & thermonuclear bursts in case of NS primaries & \\

\hline
\end{tabular}
\tablerefs{ (1) \citet{Warner03}; (2) \citet{Echevarria88}; (3)\citet{Silber92}; (4)\citet{Hynes04}; (5)\citet{Zurita03}; (6)\citet{Remillard06}; (7)\citet{Marsh94}; (8)\citet{Menou99}; (9)\citet{Lasota01}; (10)\citet{Casares91}; (11) \citet{Grindlay99}; (12)\citet{Shara05}; (13)\citet{Grindlay06}}
\footnotetext[1]{Average and standard deviation of reported EW ratios in \citet{Echevarria88}}
\footnotetext[2]{For systems where EW(H$\beta$)$>20\,\AA$. This is not a definitive test, as it is based on tens of systems. The defining characteristic of IPs is an X-ray spin period for the WD that is less than the orbital period. Our X-ray observations are too shallow to permit detections of a spin period in most cases.}
\footnotetext[3]{Reported values are average of observed EW ratios for the systems A0620-00 and V404 Cyg in quiescence and outburst, respectively.}
\label{tab:sort}
\end{center}
\end{table*}

When the magnetic field of the WD is strong enough it can lock the WD to the companion star, so that $P_{\mathrm{Spin}} = P_{\mathrm{Orb}}$, and funnel material to the magnetic pole of the WD directly from L1. These are called AM Her systems, after the proto-typical example, or Polars. 

Active Galactic LMXBs are some of the brightest X-ray sources in the sky, with examples including the brightest persistent X-ray source, Sco X-1. The ratio of X-ray to optical flux for such systems is on the order of 100 or more ($\frac{F_X}{F_{Opt}} \geq 100$). When these systems enter quiescence, the X-ray flux drops significantly more than the optical light, resulting in a flux ratio closer to order unity for neutron star systems. Quiescent LMXBs (qLMXBs) with a black hole (BH) primary are less luminous in the X-ray than NS systems, which is thought to be a result of heated material falling beyond the event horizon of a BH before radiation can escape, whereas heated infalling material on the surface of a NS can eventually radiate the energy away \citep{Narayan97,Garcia01,Hameury03,Narayan08}. Another scenario is that the energy escapes the system through jets in BH systems \citep{Fender03}. The qLMXBs we expect to find in the GBS have not gone through a recent outburst cycle; we are searching for quiescent systems, not following systems into quiescence from outburst. It is possible, therefore, that the quiescent properties of systems in the GBS are different from known qLMXBs.

Some counterparts to GBS sources are visible in the Optical Gravitational Lensing Experiment data \citep{Udalski12}. Crowding in these fields is high enough that sources should be treated with care, as the odds of a chance alignment of the X-ray position with an unrelated variable star are non-negligible. 

We provide a rough rubric to differentiate between these compact binary systems in Table \ref{tab:sort}. Values are drawn from existing literature on each category, and are based on a limited number of well observed CVs and only one BH qLMXB and one NS qLMXB. Though they are representative of the characteristics of similar systems, they should be taken with care. Often, the optical flux in existing literature refers to $V$ band measurements. Our photometry consists of Sloan $r'$ observations, and will henceforth report flux values as $F_{r'}$ rather than $F_{Opt}$. The difference in these broad band measurements depends on the spectral energy distribution of the source, but is small compared to the difference with the X-ray flux. 

In this paper we examine the first five accreting binaries identified in the GBS by means of long-slit optical spectroscopy and optical photometry.

\begin{figure*}[h!]
\begin{center}
  \includegraphics[width=0.3\textwidth, angle=0]{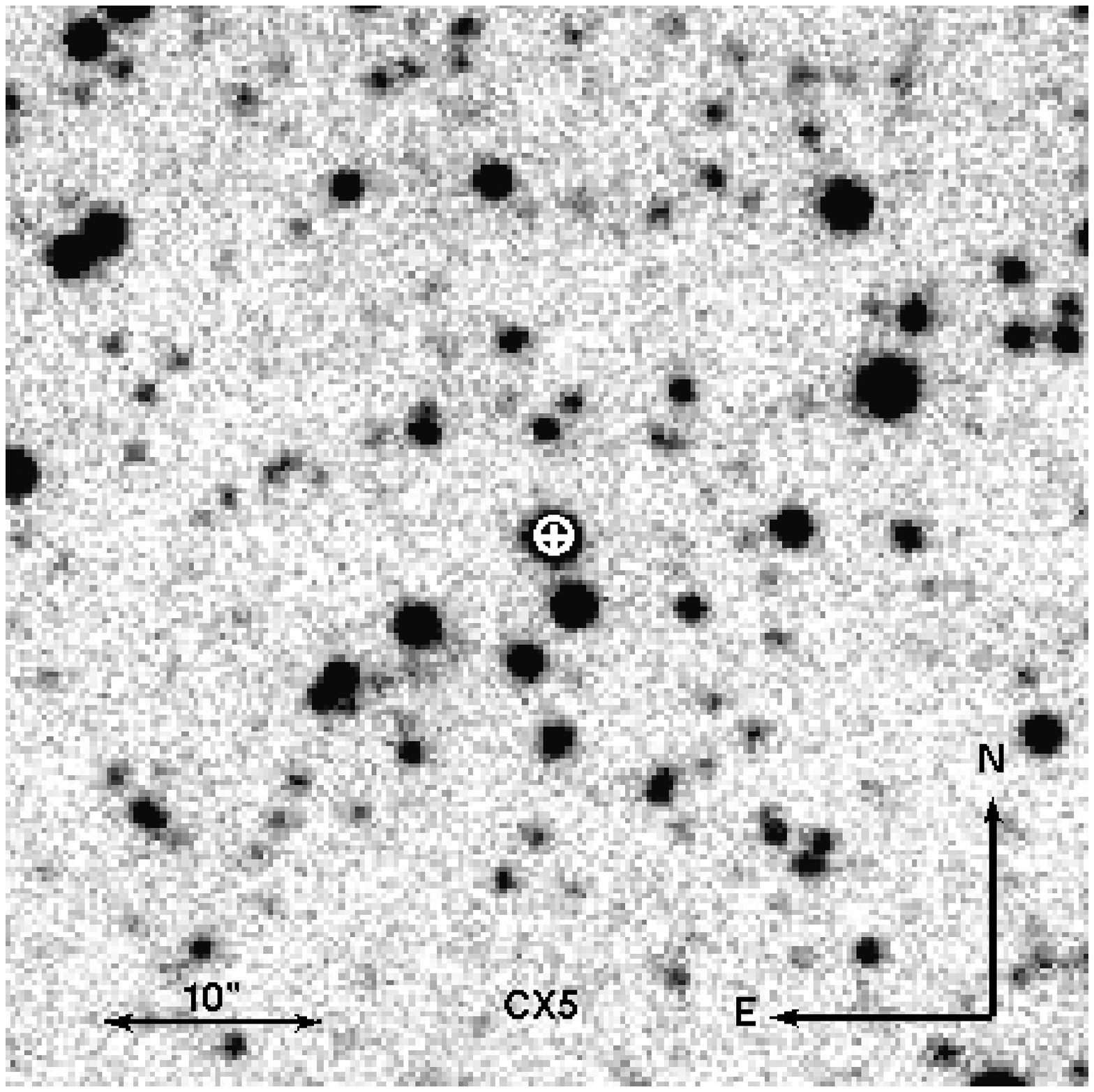}
  \includegraphics[width=0.3\textwidth, angle=0]{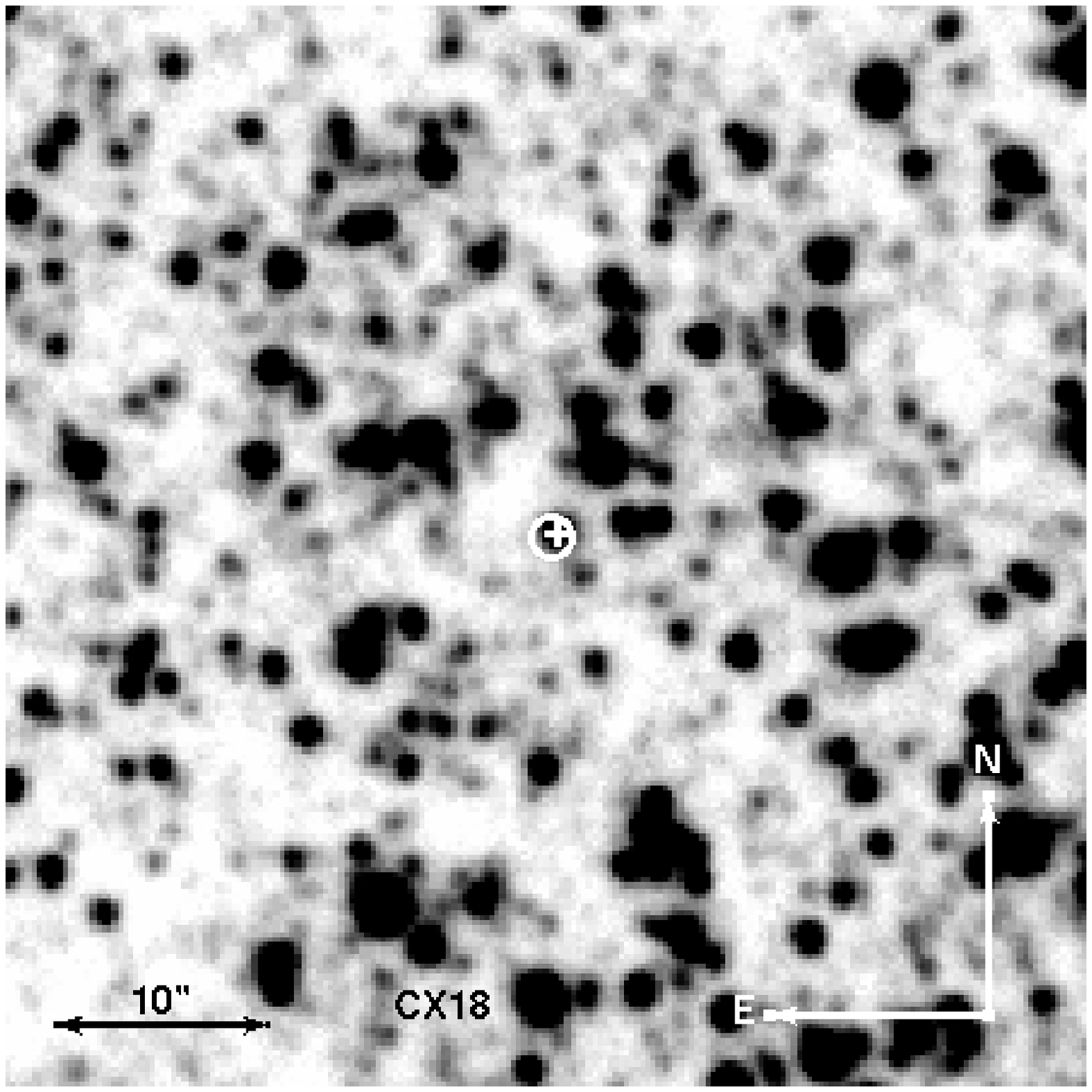}\\
  \includegraphics[width=0.3\textwidth, angle=0]{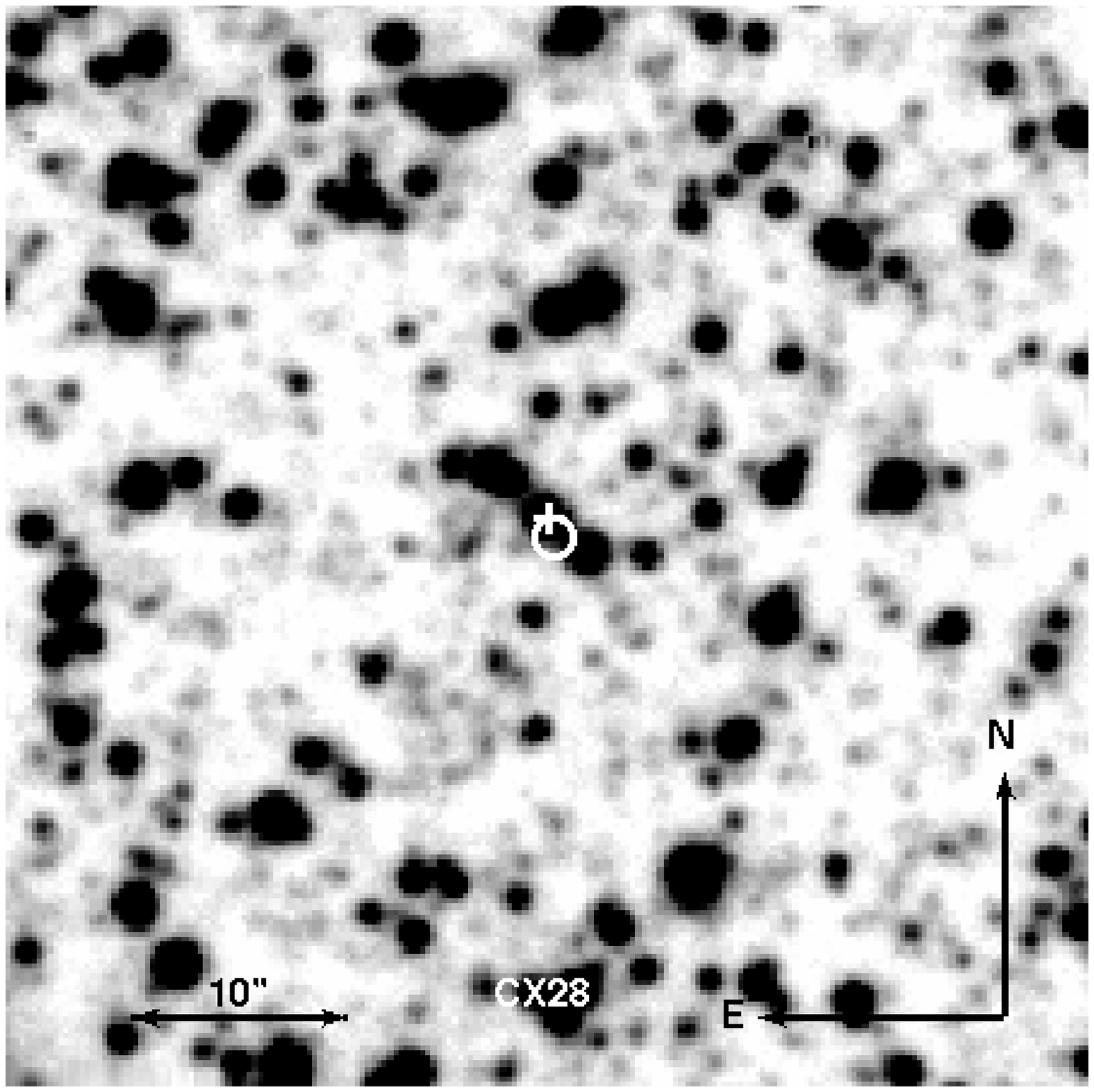}
  \includegraphics[width=0.3\textwidth, angle=0]{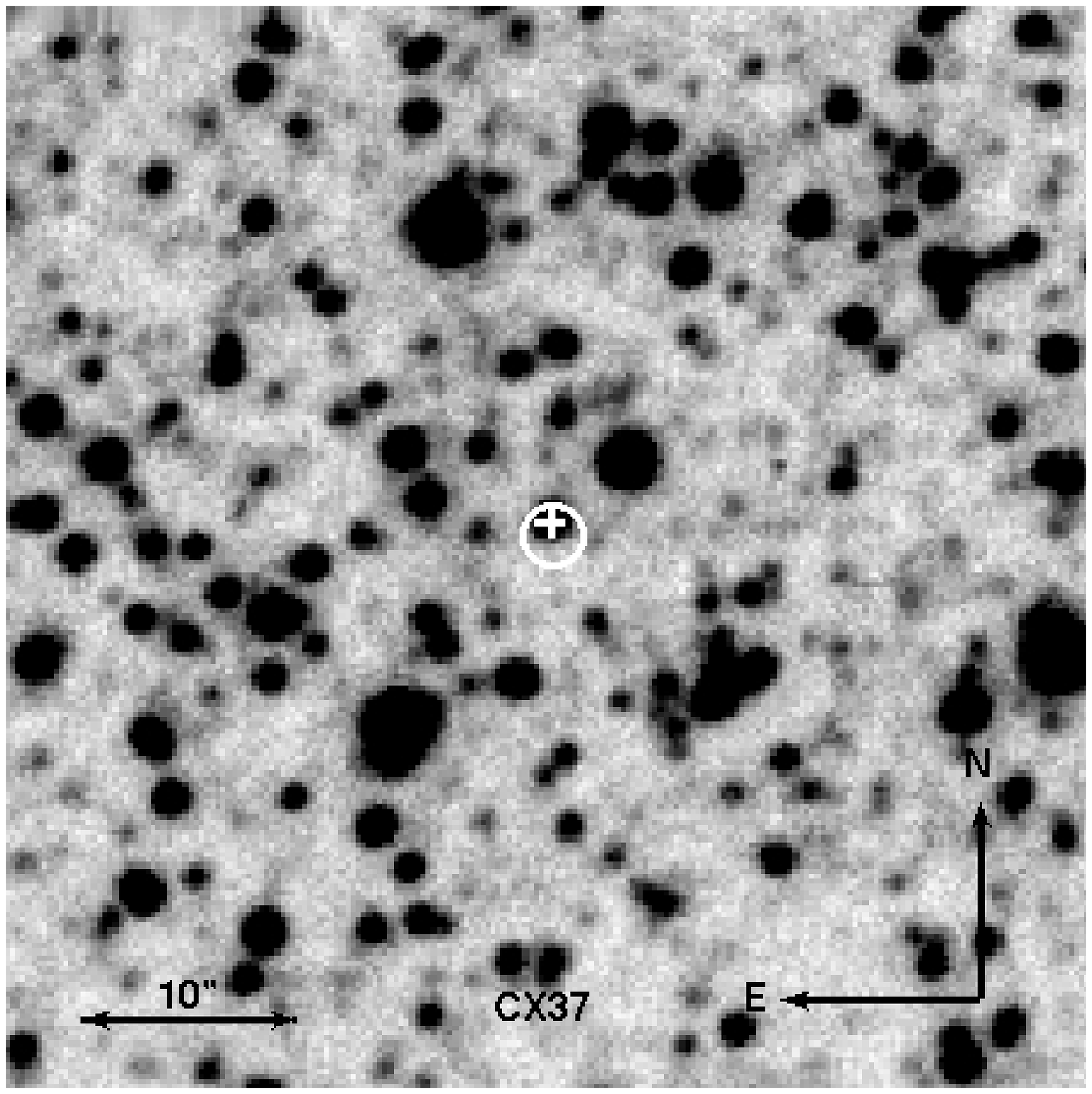}
  \includegraphics[width=0.3\textwidth, angle=0]{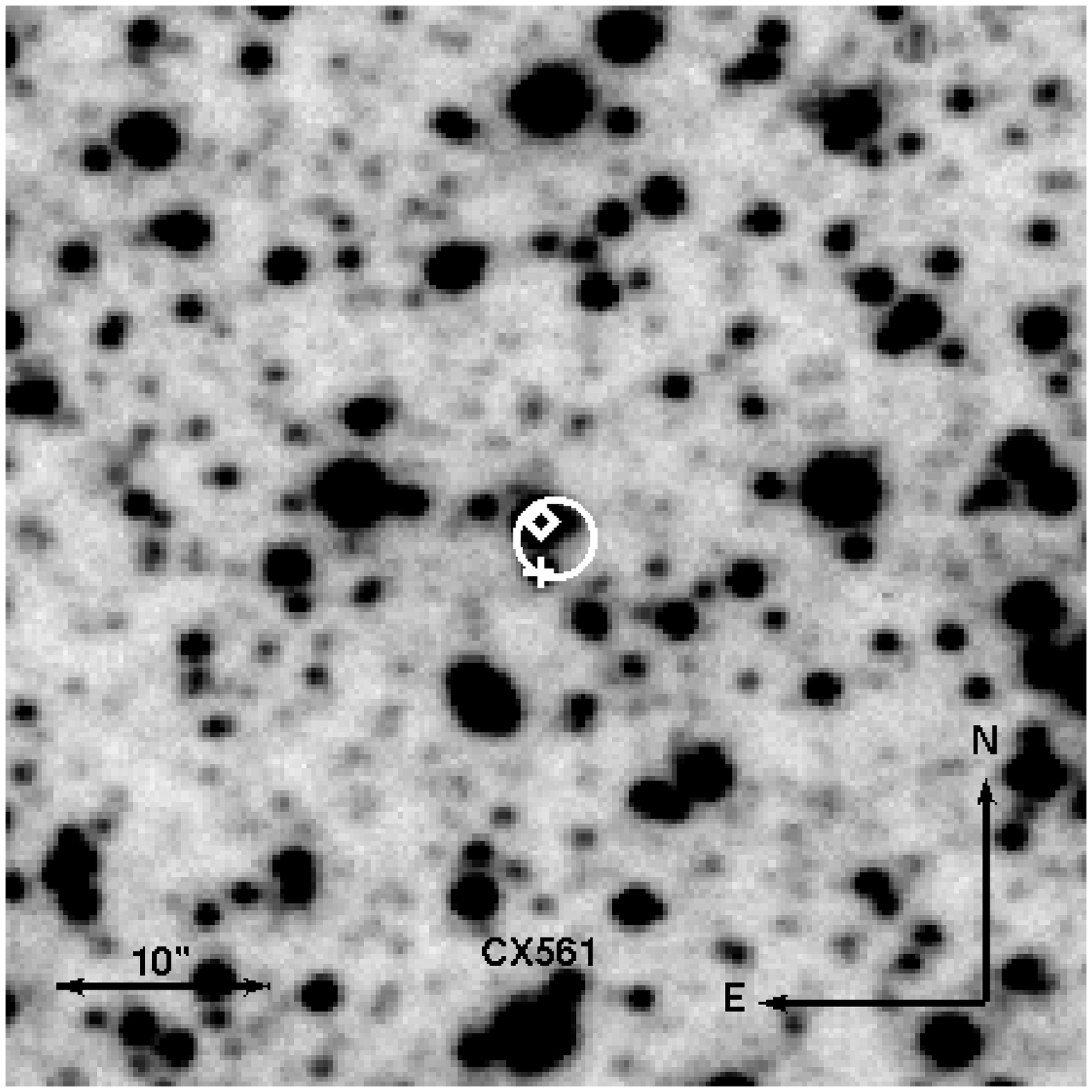}

\caption{Finder charts for the 5 listed GBS sources. The 3-sigma X-ray position is plotted with a white oval, while the optical companion is marked with a white cross a) CX5 b) CX18 c) CX28 d) CX37 e) CX561, with a nearby eclipsing binary system marked with a white diamond. In all charts, North is up and East is left.}
\label{fig:finders}
\end{center}
\end{figure*}

\section{Data Reduction}

\subsection{Spectroscopy}
We targeted the five objects presented in this paper during our spectroscopic campaigns to identify and classify the optical counterparts to GBS X-ray sources. Data were acquired between July 8th, 2010 and July 11th, 2010 under program 085.D-0441(C) with the New Technology Telescope (NTT) equipped with the ESO Faint Object Spectrograph and Camera (EFOSC2). These 5 sources are selected from among the targets of the 2010 NTT observations as the only sources showing strong emission lines in the spectra. Results on the other sources will be presented elsewhere. The observations were performed using grism \#13 and a 1'' wide slit that provided an instrumental spectral resolution of $\sim 17$\, \AA\ (FWHM) in the $\lambda\lambda = 3700-9200$\,\AA\ wavelength range. Integration times ranged between 400\,s and 900\,s. HeAr arc lamp and flat field exposures  were taken after each target observation. Additionally, on April 2nd, 2011 two consecutive 875\,s spectra of the optical counterpart to CX28 were obtained with the Visible Multi Object Spectrograph (VIMOS; LeFevre et al. 2003) mounted on the ESO Very Large Telescope under program 085.D-0441(A). The MR red grism and a 1'' wide slit were used, achieving a spectral resolution of $10$\,\AA\  FWHM and a wavelength range of $\lambda\lambda = 4600-10000$\,\AA.

The EFOSC2 data set was bias and flat-field corrected with standard IRAF\footnote{IRAF is distributed by the National Optical Astronomy Observatory, which is operated by the Association of Universities for Research in Astronomy (AURA) under cooperative agreement with the National Science Foundation.} routines. The spectra were extracted with the IRAF {\sc kpnoslit} package. The pixel-to-wavelength calibration was derived from cubic spline fits to HeAr arc lines. The root-mean square deviation of the fit was $< 0.1$\,\AA. Checks for the stability of the wavelength calibration were made using the atmospheric [O I] 5577.34 and $6300.3$\,\AA\ sky lines. In this way we estimated an accuracy in the wavelength calibration of $<0.5$\,\AA. The VIMOS spectra of CX28 were reduced with the VIMOS ESO pipeline version 2.6.2 (Izzo et al. 2004) and extracted using IRAF (see Torres et al. 2013 for details).

\begin{figure*}[h!]
\begin{center}
  \includegraphics[width=0.3\textwidth, angle=0]{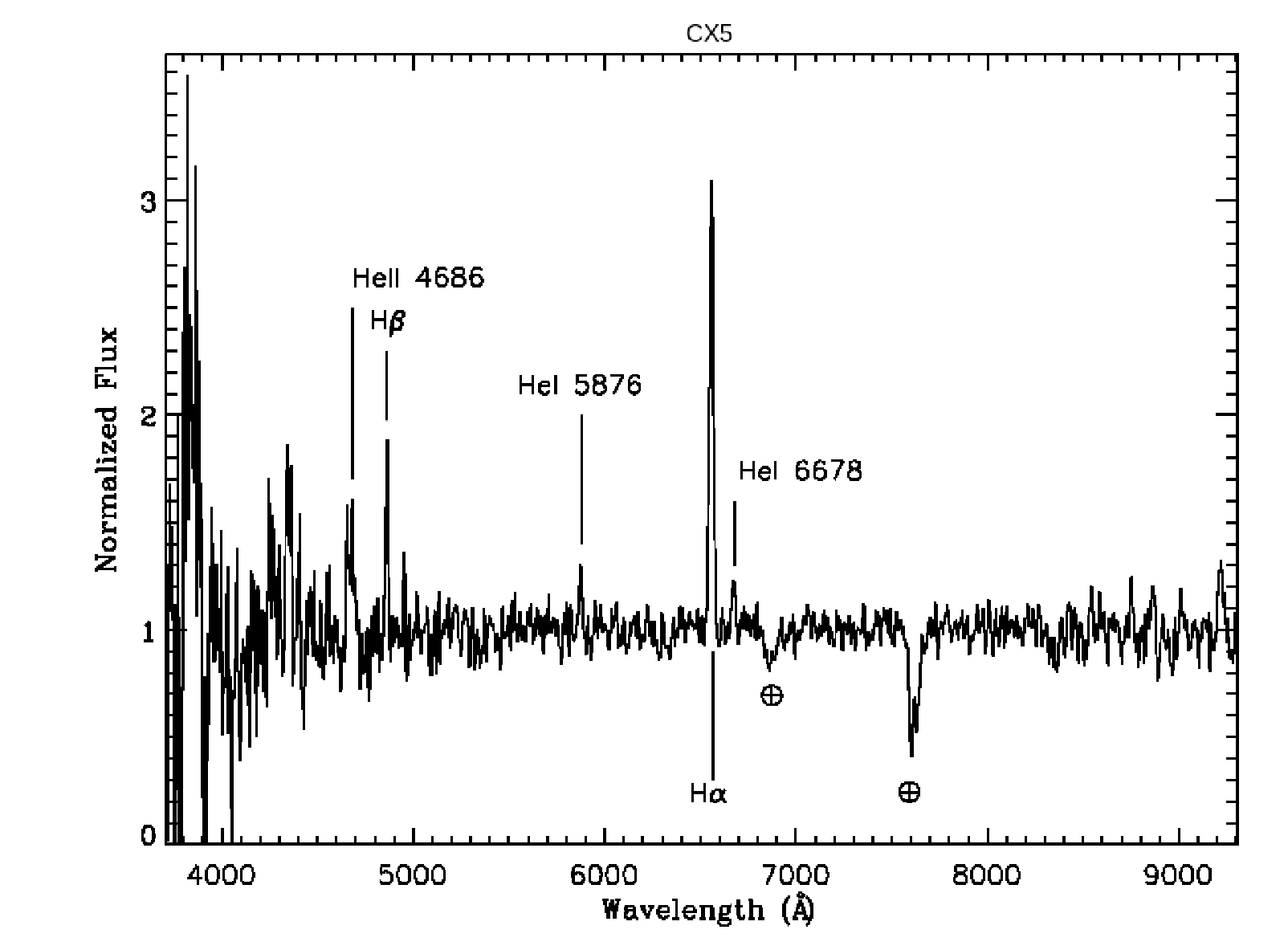}
  \includegraphics[width=0.3\textwidth, angle=0]{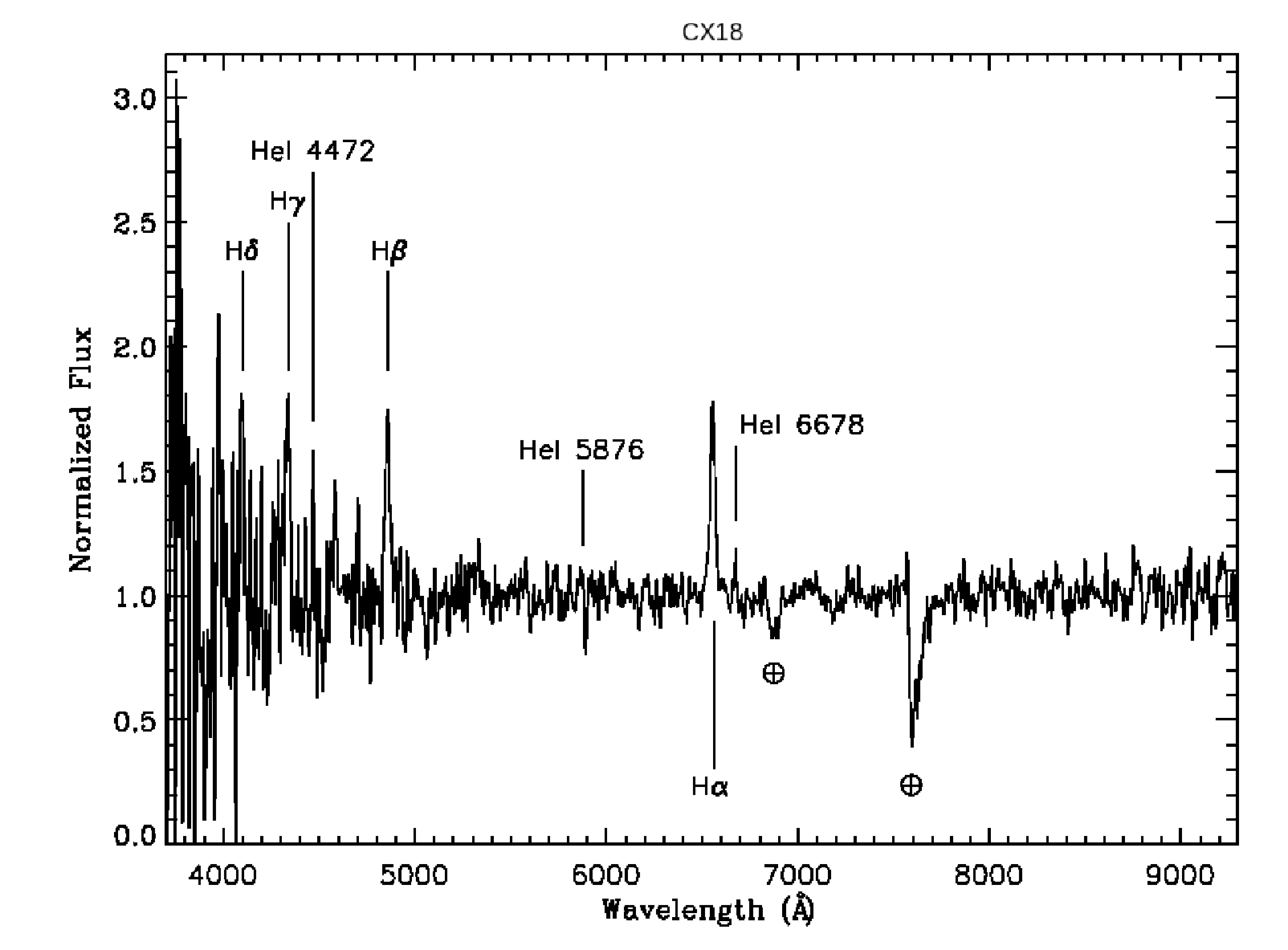}
 \includegraphics[width=0.22\textwidth, angle=90]{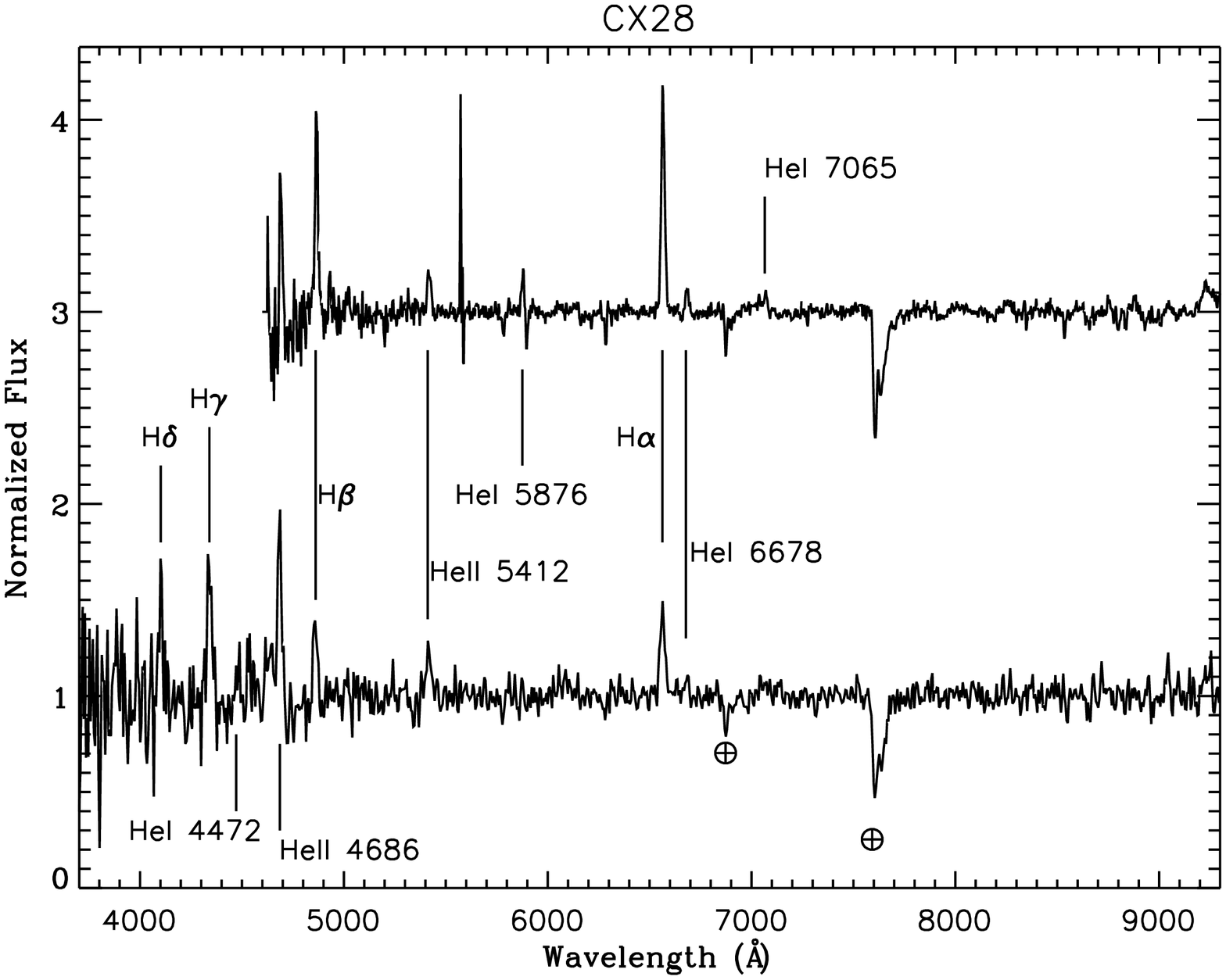}\\
 \includegraphics[width=0.3\textwidth, angle=0]{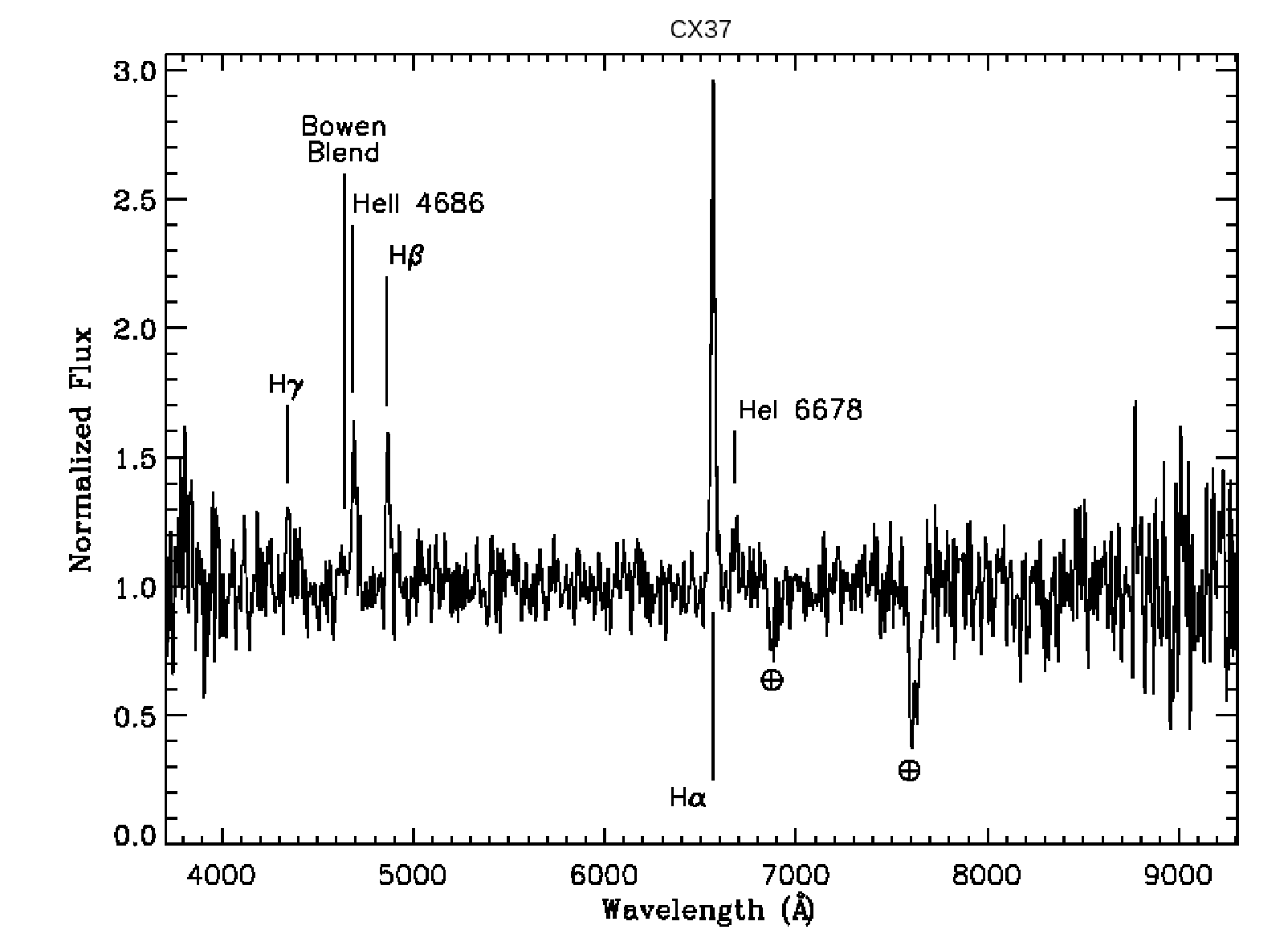}
 \includegraphics[width=0.3\textwidth, angle=0]{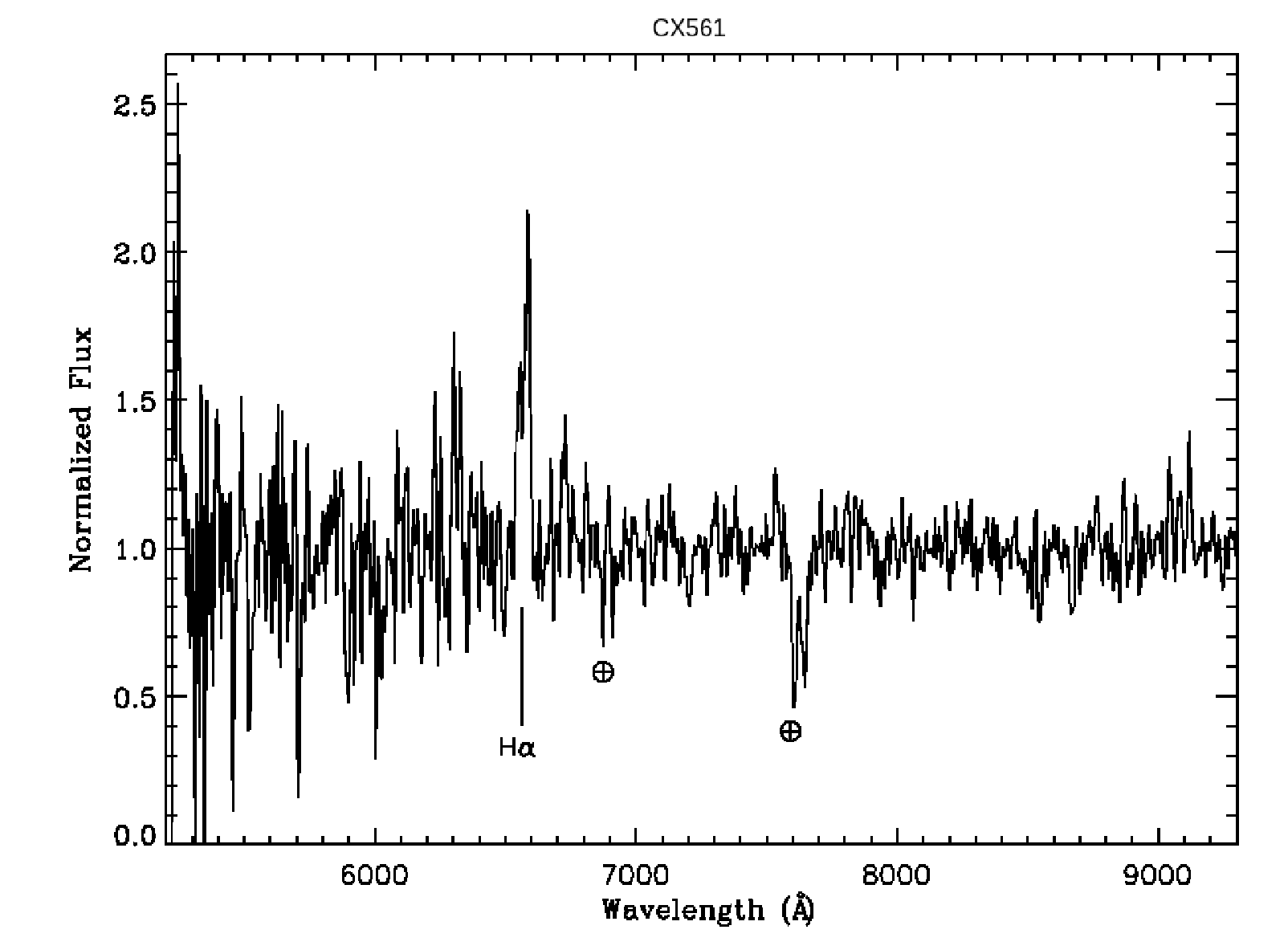}

\caption{a) 2 epochs of the NTT spectrum for CX5. b) 2 epochs of the NTT spectrum for CX18. c) The bottom epoch is from NTT observations of CX28, while the top comes from the VIMOS instrument on the VLT.  d) The NTT spectrum for CX37.  e)The NTT spectrum for CX561.}
\label{fig:spectra}
\end{center}
\end{figure*}

\subsection{Optical Photometry}

\subsubsection{Blanco Photometry}
We acquired 8 nights of photometry, from July 12th-18th 2010, with the Blanco 4.0 meter telescope at the Cerro Tololo Inter-American Observatory (CTIO). Using the Mosaic-II instrument, we observed the 9 square degree area containing two thirds of the X-ray sources identified by the GBS \citep{jon11}. Multiple Sloan $r'$ band exposures with an integration time of 120\,s of 45 overlapping fields were taken to cover the area. Typical seeing for the run was around 1''. The order in which the fields were cycled was randomized to minimize aliasing caused by regular sampling. The data were reduced via the NOAO Mosaic Pipeline \citep{Shaw09}, which also added a world coordinate system to the images. 

The NOAO pipeline searches for instrumental artifacts in the image, corrects for cross talk between CCDs, applies a pupil ghost correction for light reflecting from the filter to the back surface of the corrector then back through the filter, applies bias and flat field corrections, and calibrates WCS for each image based on USNO-B1 stars in the field. Dark current calibrations are unneccesary. A detailed explanation of each procedure can be found in chapter 2 of the NOAO Data Handbook \citep{Shaw09}.

Photometry on the 5 sources with NTT spectra showing emission lines was done using Alard's image subtraction routine, ISIS, described in detail in \citet{Alard98} and \citet{Alard00}. ISIS works by using a reference image which it then convolves with a kernel in an effort to match a subsequent image of the same field. The subsequent image is then subtracted from the convolved reference image. Stars which do not vary in magnitude should subtract cleanly, so the subtracted image is clear of non-variable objects. Therefore, any residual flux is due to an inherent change in brightness of a source. To perform photometry on the subtracted image, the point spread function (PSF) for each image is scaled by $\chi^{2}$ minimization to represent the change in flux from the reference image. The error bars are calculated through $\chi^{2}$ minimization as well. Systematic errors in determining the PSF across each image are not accounted for, however. In most cases, these errors are quite small as demonstrated by cleanly subtracted frames. In order to save computation time, small cutouts of the full Mosaic images, $401 \times 401$ pixels or $104'' \times 104''$, were taken around each object for processing.

\begin{figure*}[h!]
\begin{center}

  \includegraphics[width=0.35\textwidth, angle=90]{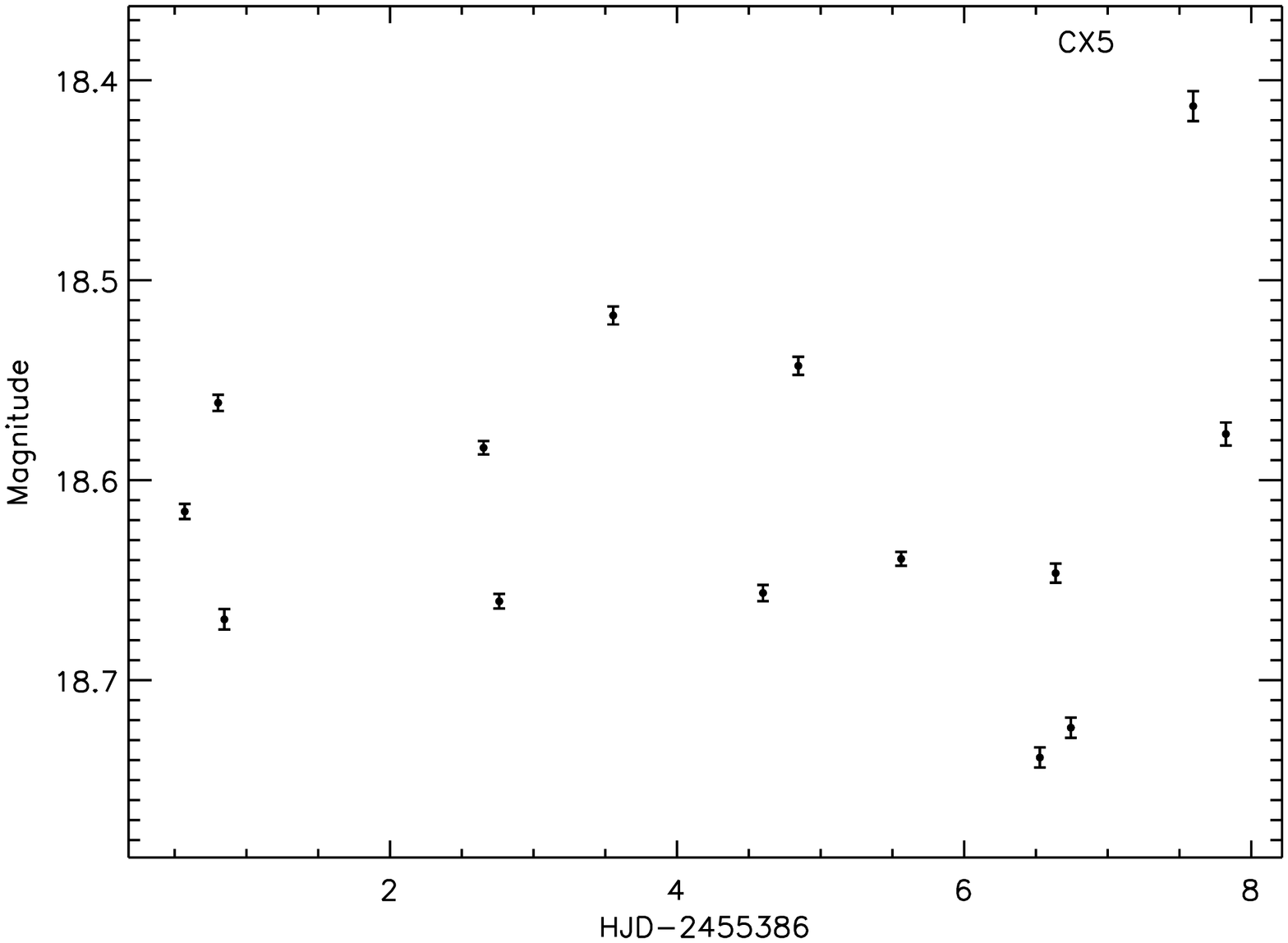}
  \includegraphics[width=0.35\textwidth, angle=90]{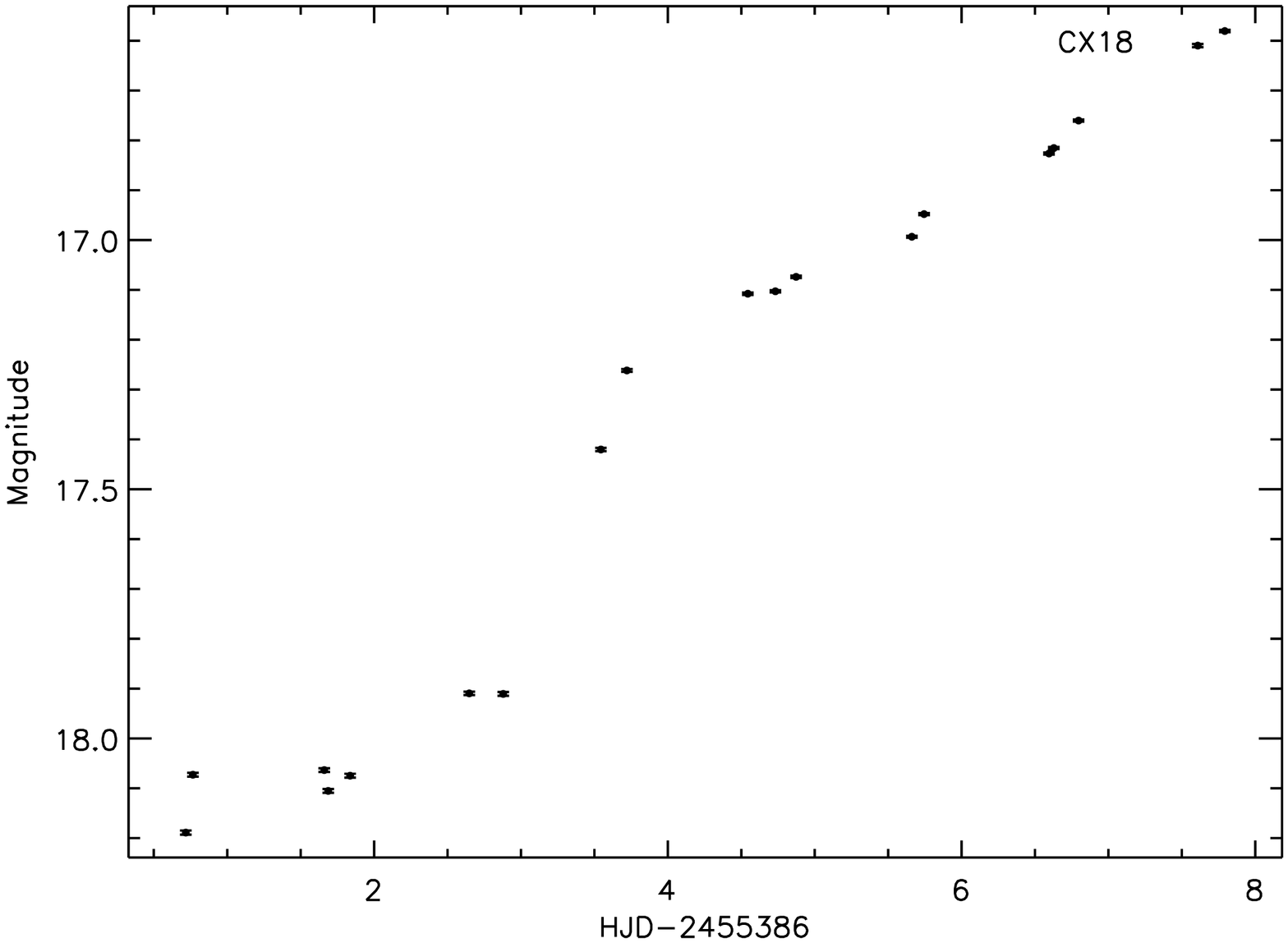}\\
  \includegraphics[width=0.35\textwidth, angle=90]{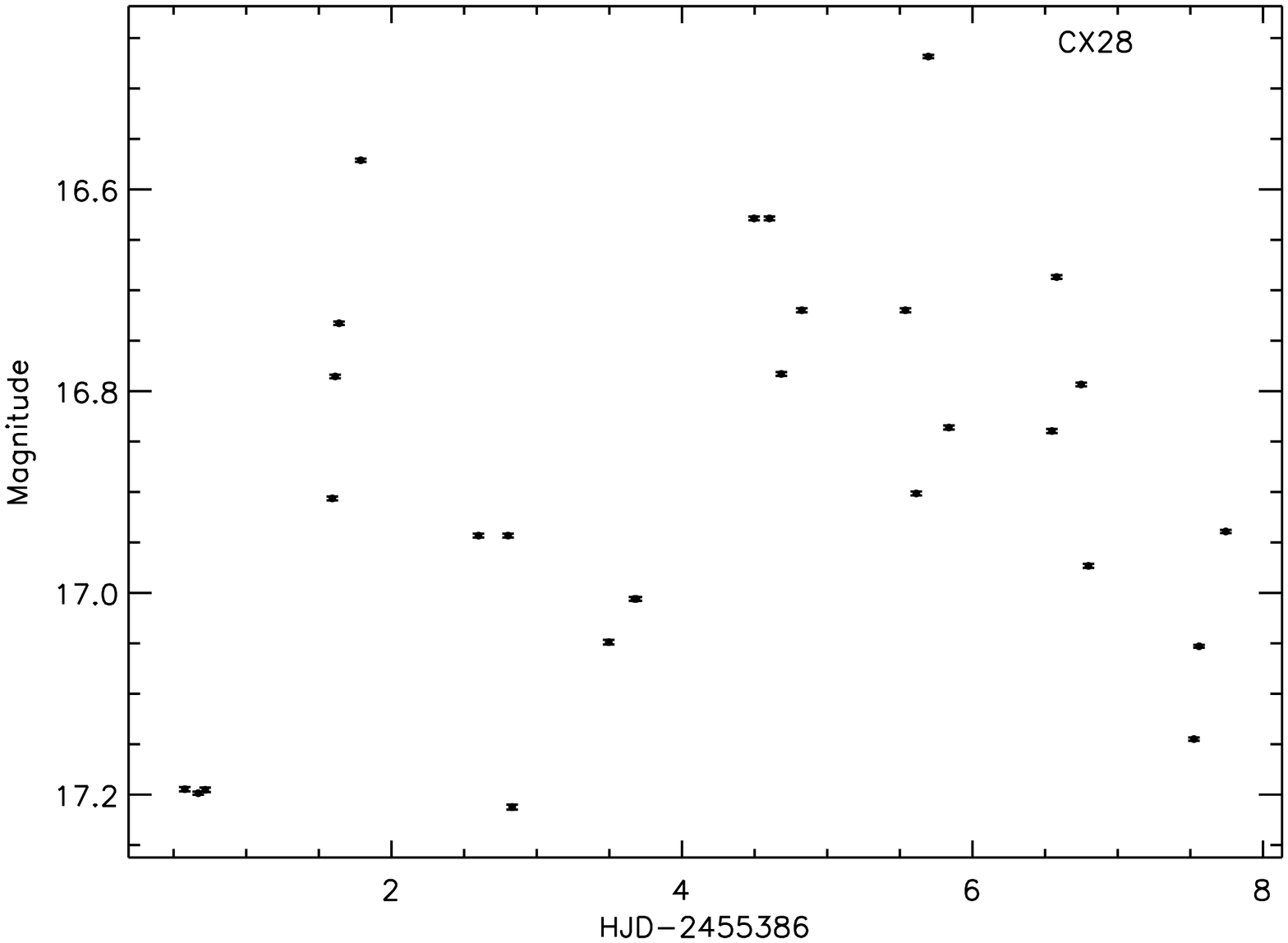}
  \includegraphics[width=0.35\textwidth, angle=90]{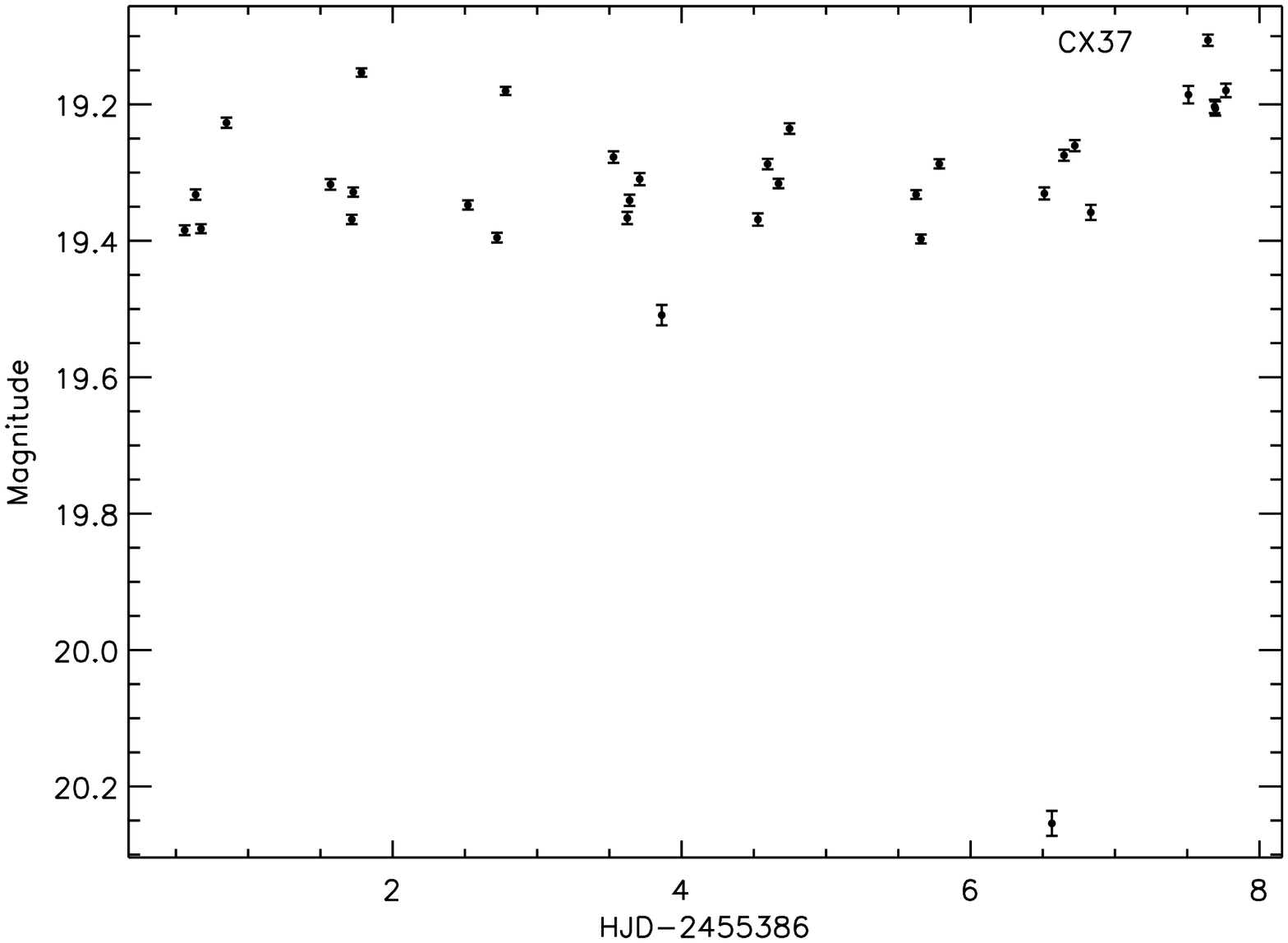}
\caption{A) Lightcurve of CX5 from Mosaic-II data. B) Lightcurve of CX18 from Mosaic-II data, showing a rise of 1.5 magnitudes over 6 days.  C) Blanco Mosaic-II lightcurve of CX28. D) Blanco Mosaic-II lightcurve of CX37, showing an eclipse 4.39 times the standard deviation of the lightcurve below the median magnitude.}
\label{fig:lc}
\end{center}
\end{figure*}

Once a lightcurve was in hand, if a source showed significant variability, periodograms were created using the Lomb-Scargle statistic in an effort to search for periodicities. Since ellipsoidal variations have 2 maxima and minima in a single orbital period, we also check periods twice as long as prominent peaks on a periodogram. We also consider both aliases and harmonics, as higher harmonics can sometimes show up at a higher power than the fundamental frequency.

At present, we lack photometric standard observations for these sources in $r'$, so all apparent magnitudes cited here are scaled to nearby stars in the USNO-B1 catalog and are to be used with caution until secondary standards are established for all Mosaic fields. The magnitude scaling, which is a pipeline calibration product, carries an estimated uncertainty of $\pm0.5$ magnitudes for each source. This is quite adequate for estimating X-ray to optical flux ratios.

\subsubsection{SMARTS Photometry}

We also used the SMARTS Consortium's 1.3m at CTIO to gather further optical data for CX18 and CX37 with the ANDICAM instrument. Exposures were taken with a 250\,s integration time in the $R$-band filter. Typical seeing was $1''$.  These data were reduced via pipeline, which added overscan corrections, bias corrections, and applied dark current exposures and dome flats taken by the queue observer.  Photometry was performed with ISIS as described above.

\subsubsection{Swope Photometry}

For objects showing suspected rapid variability, we observed from June 21st, 2011 through June 27th, 2011 at the 1.0 meter Henrietta Swope telescope with the SITe\#3 CCD at Las Campanas Observatory. We observed in the Gunn $r$ filter, and tracked the objects to confirm periods found with Mosaic-II data and provide more complete phase coverage of each source. Exposure times varied from 2-5 minutes depending on the brightness of the counterpart. Typical seeing was $\sim 1.3''$.  These data were reduced using standard IRAF procedures in the CCDRED package with bias and flat field frames taken each night; photometry was performed with ISIS as above.

\subsubsection{Extinction Corrections}

To correct for extinction due to dust, in this paper we use $A_{\rm K}$ and $E(B-V)$ values from Gonzalez et al (2012) and the extinction law in \citet{Cardelli89}. We transform these to $r'$ using filter property values in Schlegel et al (1998). To determine X-ray absorption, we use the relation $N_{H} = 5.8 \, \times \, 10^{21} \, {\rm cm}^{-2} \, E(B-V)$ found by \citet{Bohlin78}. \citet{Predehl95} find a value of $N_{H} = 1.79 \, \times \, 10^{21} \, {\rm cm}^{-2} \, A_{V}$, which, using the extinction law in \citet{Cardelli89}, does not differ substantially from the findings of \citet{Bohlin78}. For CX5, where there are previously reported X-ray spectral properties, we use those.  For other suspected IPs, we assume a model spectra of Brehmstrahlung radiation with $kT = 25\,{\rm keV}$, while for the remaining sources we assume a power law spectral shape of $\Gamma = 2$ \citep{jon11}. It is important to note that these extinction values are upper limits along the line of sight to the red clump stars used in \citet{Gonzalez12} rather than extinctions based on each object's actual reddening. Since extinction has a larger effect on optical wavelengths than X-rays, using these extinction values provides a lower limit to the X-ray to optical flux ratio. This could be quite extreme, since the CVs are likely in the foreground.

\begin{figure*}[h!]
\begin{center}

  \includegraphics[width=0.35\textwidth, angle=90]{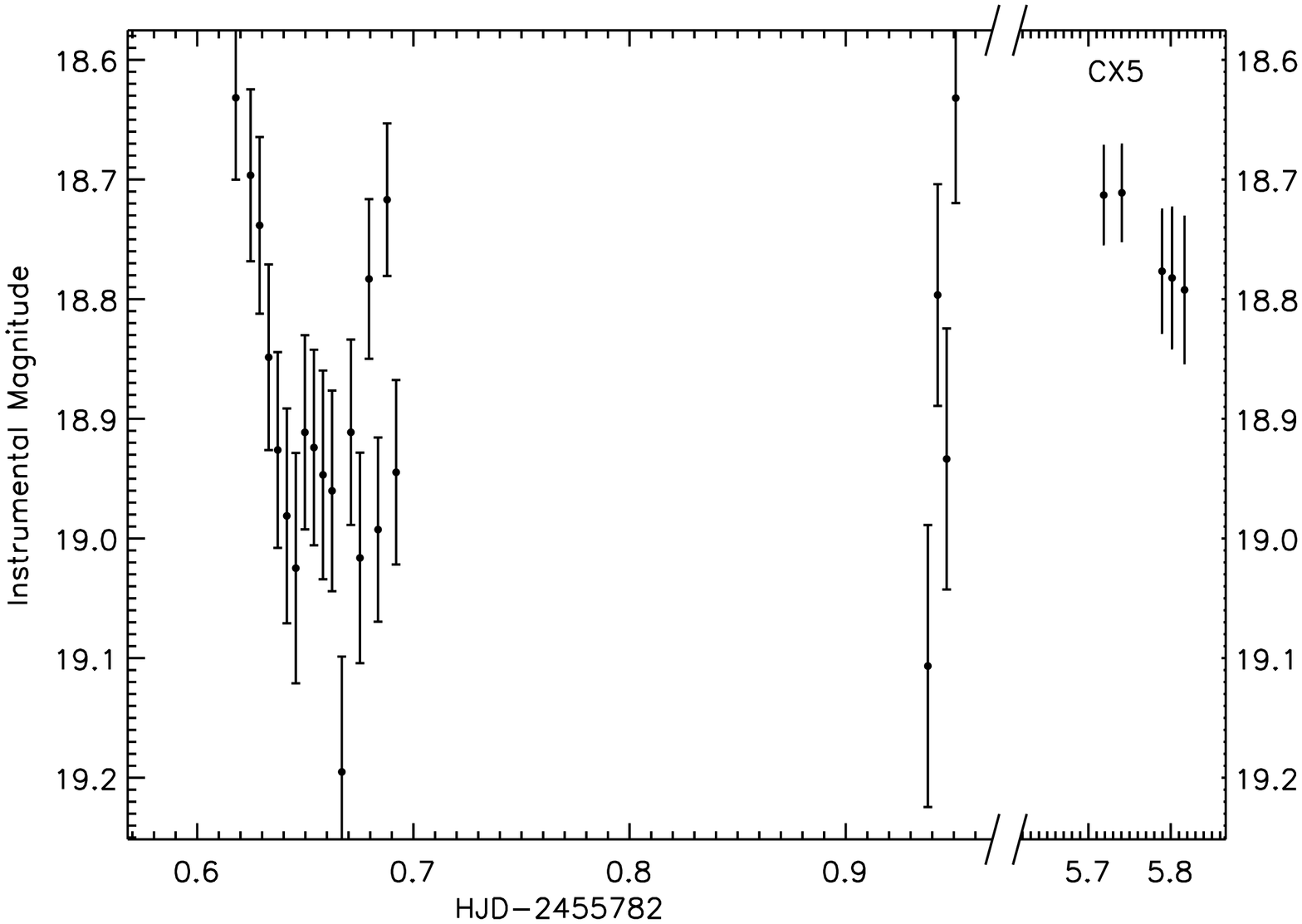}
  \includegraphics[width=0.35\textwidth, angle=90]{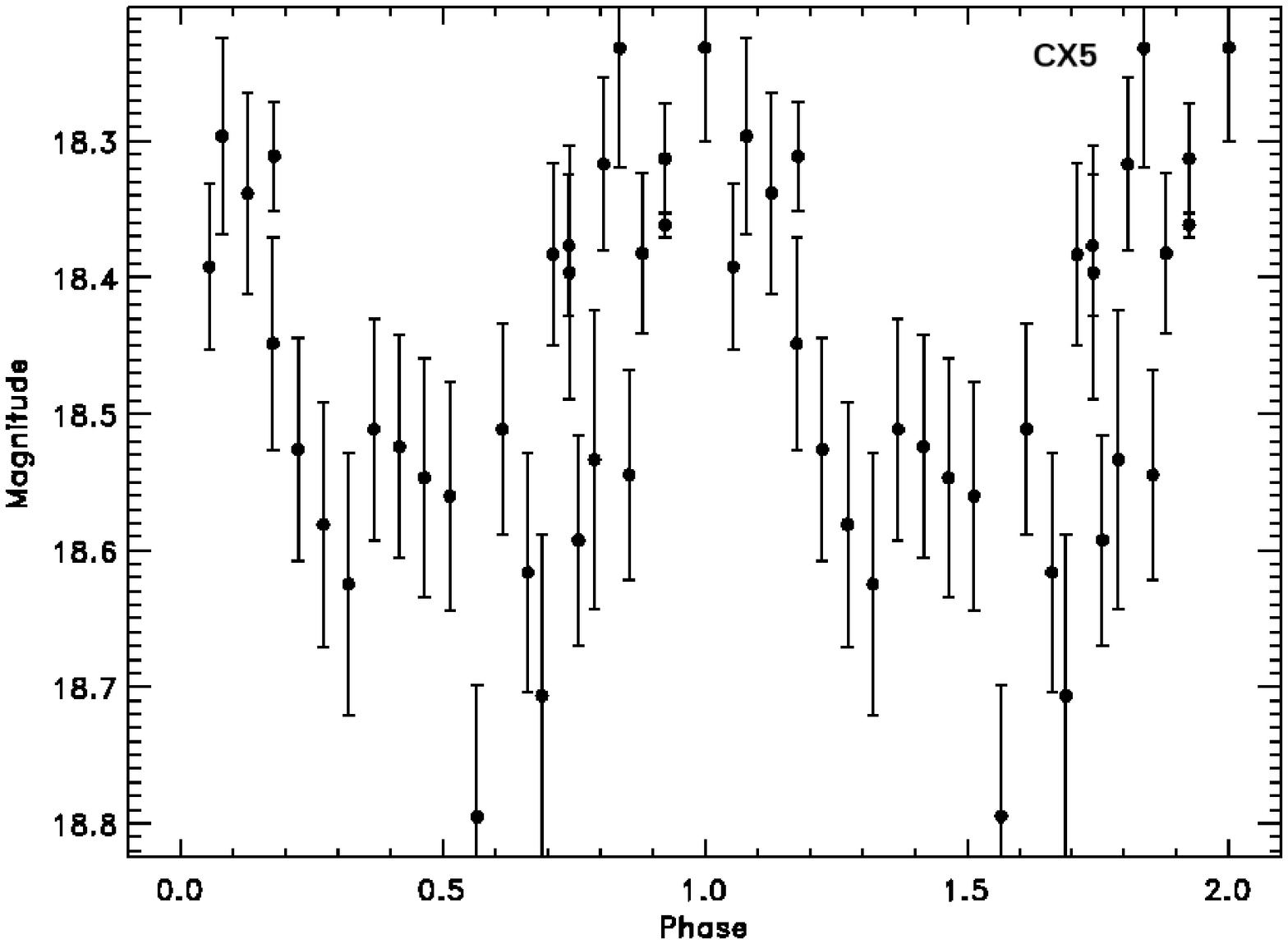}\\
  \includegraphics[width=0.35\textwidth, angle=90]{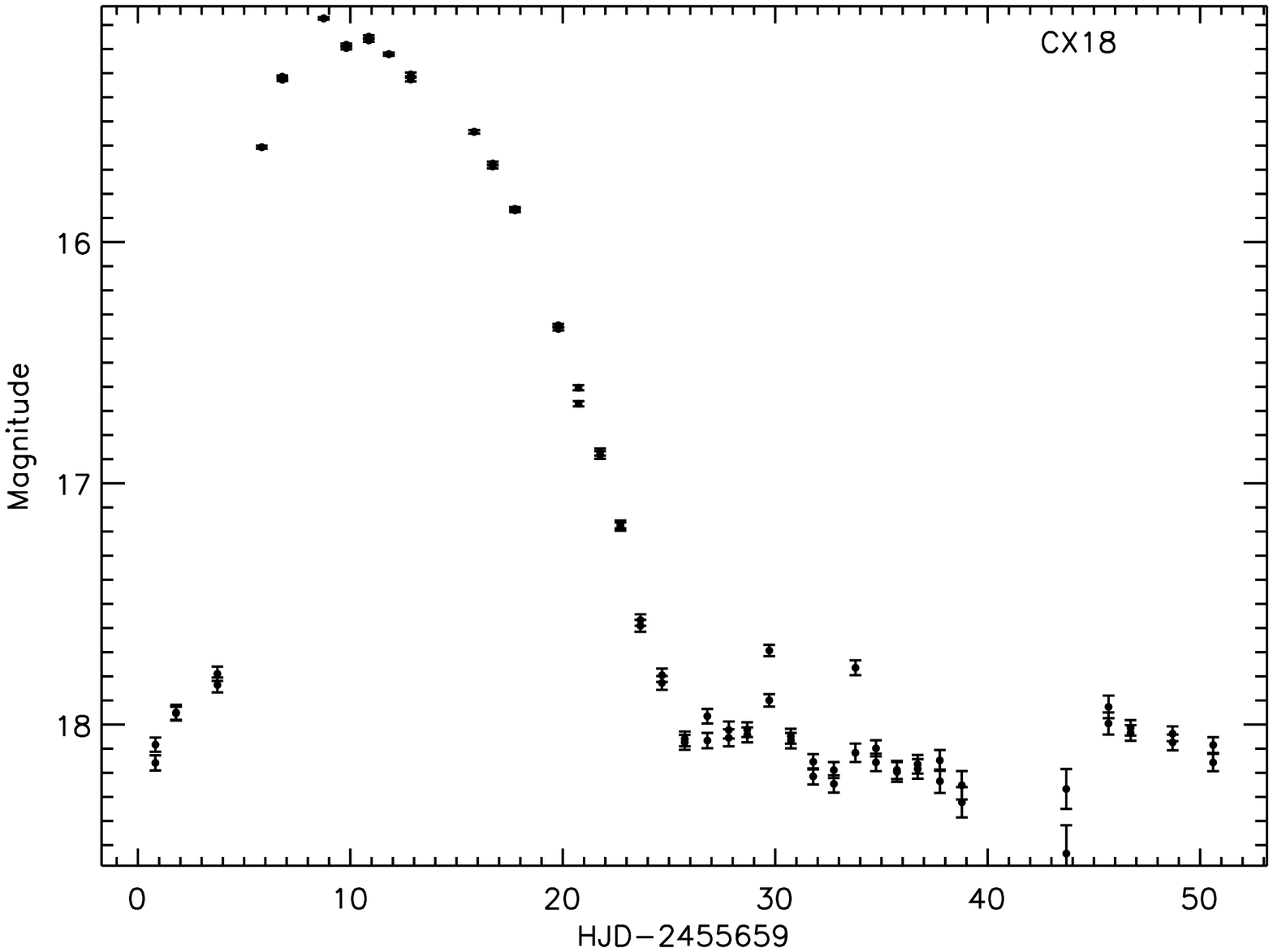}
  \includegraphics[width=0.35\textwidth, angle=90]{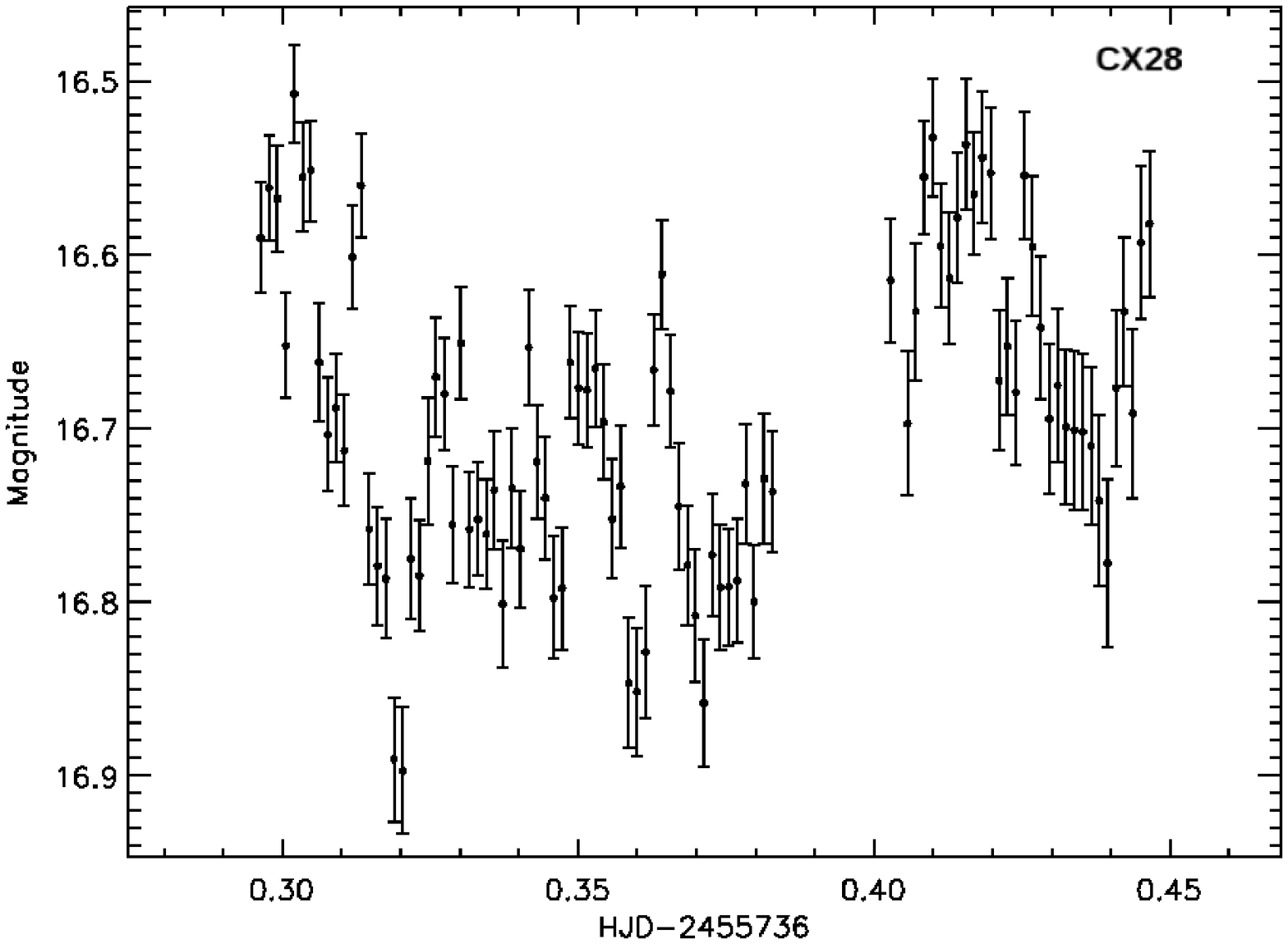}

  \caption{a) Lightcurve of CX5 from the Swope 1.0 meter data taken in the summer of 2011. 
b) Swope lightcurve of CX5 folded on a 125 minute period with an arbitrary ephemeris. 
c) Data collected with the SMARTS 1.3m Andicam in the spring of 2011 confirms that CX18 is undergoing dwarf nova outbursts with a fast recurrence time. 
d) CX28 Lightcurve from Henrietta Swope telescope at LCO. }

\label{fig:lcfollowup}
\end{center}
\end{figure*}

\section{Results}
 Fig. \ref{fig:finders} shows a cutout of the Blanco image around each X-ray source. Table \ref{tab:lines} presents the emission line parameters for each source, while the spectra themselves are in Fig. \ref{fig:spectra}.  Lightcurves of Mosaic-II data are shown in Fig. \ref{fig:lc}, while follow up lightcurves are presented in Fig. \ref{fig:lcfollowup}. Of the 5 objects presented, 1 shows an eclipse, 1 shows a possible 2 hour period, 2 display aperiodic flickering, and 1 is not variable.

\subsection{CXOGBS J174009.1-284725 (CX5)}

 This source is also known as AXJ1740.1-2847 and shows an X-ray period of 729s (Sakano et al 2000). In an IP interpretation, this pulse period is the spin of the WD. 
Sakano et al (2000) fit the X-ray spectrum of CX5 to a power law of index $\Gamma = 0.7\pm 0.6$ with $N_{H} = 2.5^{+2.9}_{-1.8} \times 10^{22} \, {\rm cm^{-2}}$. \citet{Kaur10} find $N_{H} = 1.0 \pm 0.2 \times 10^{22} \, {\rm cm^{-2}}$ with $\Gamma = 0.5\pm 0.1$.  \citet{Kaur10} also report the presence of Fe emission. Its optical spectrum shows He II emission lines, which rules out a qLMXB interpretation. The ratio of equivalent widths between He II $\lambda$ 4686 and H$\beta$ is 2.3, which is suggestive of a magnetic CV. Using the value in \citet{Kaur10} for $N_{H}$ to calculate extinction and absorption, we find the ratio of X-ray to optical flux is $\frac{F_{x}}{F_{r'}} \sim 0.4$ which is consistent with an IP, CV, or qLMXB.  Assuming a distance of 1 kpc yields $L_X = 3 \times 10^{32}\, {\rm ergs\, s^{-1}}$, which is within errors of previously reported values for this source.

This source shows variability of 0.3 magnitudes in our Mosaic-II data, the lightcurves of which are shown in Fig. \ref{fig:lc}.
 The Mosaic-II observations for this source do not show a well defined periodicity. We obtained 6 hours of intermittent photometry with the Henrietta Swope Telescope at Las Campanas Observatory from June 22nd, 2011 and turned to this source again on June 27th, 2011. We see evidence of a 125 minute period in the optical (see Fig. \ref{fig:lcfollowup}). The Mosaic-II observations do not show evidence of this period. Using $P_{spin} \sim 0.1\,P_{Orb}$ as the apparent peak of the distribution of known IPs \citep{Scaringi10}, one expects a 729\,s pulse period to result in a roughly 2 hour orbital period, lending credibility to identifying the 125 minute period as the orbital period. The previously detected X-ray period of $729\,s$, hard X-ray spectrum with Fe emission, strong He II emission, and possible 125 minute orbital period ($\sim 10 \times P_{spin}$) point to CX5 being an IP when taken together.

\subsection{CXOGBS J173935.7-272935 (CX18)}

This object is noted as a variable star in \citet{Terzan91}. The optical spectrum of CX18 shows He I and Balmer emission. 
In the Mosaic-II data, this source brightens by $1.6$ magnitudes over the course of 6 days. Follow up data from SMARTS, which can be seen in Fig. \ref{fig:lc}, clearly shows a similar rise of 3 magnitudes followed by an exponential decay and a flat quiescent state after the outburst. In two observations spaced out by several months, we caught two of these outbursts, suggesting a high recurrence rate. This source appears to be a CV undergoing Dwarf Nova (DN) outbursts, which is in agreement with \citet{Udalski12}, who find DN outbursts with a recurrence time around 100 days in OGLE data. 

Using the $E(B-V)$ values from \citet{Gonzalez12} yields an X-ray to optical flux ratio of order $\frac{1}{10}$, which is consistent with previous observations of the ratio of X-ray to optical light of quiescent DNe \citep{Verbunt97}. Most DNe are substantially closer than bulge distance, but even assuming that no absorption occurs, the unabsorbed ratio of X-ray to optical flux of order unity is within the range of ordinary CVs. DN outbursts can occur in both CVs and in IPs with sufficiently large disks. Many IPs show strong HeII emission in their optical spectra \citep{Edmonds99}. CX18 does not. The lack of He II lines combined with the comparatively strong X-ray emission indicates that this is a quiescent system at the time of the spectroscopy and argues against an IP interpretation.

\begin{figure*}[h!]
\begin{center}

  \includegraphics[width=0.25\textwidth, angle=90]{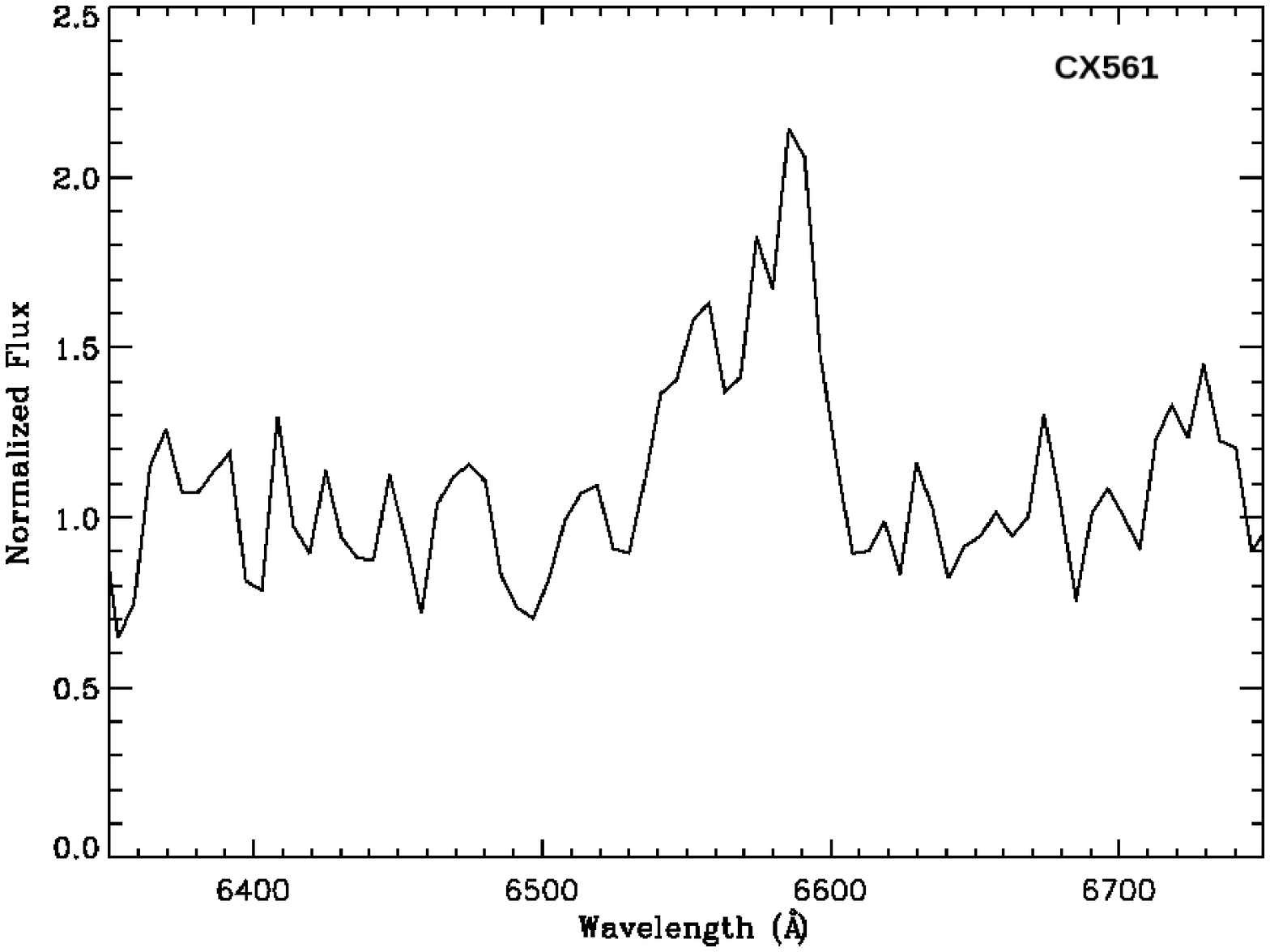}
\caption{Zoom in on the double peaked H$\alpha$ line in the NTT spectrum of CX561, peaks separated by 30 \AA or $1420 \,{\rm km\,s^{-1}}$, which implies a disk velocity of $710 \,{\rm km\,s^{-1}}$}
\label{fig:CX561}
\end{center}
\end{figure*}

\subsection{CXOGBS J173946.9-271809 (CX28)}

The optical spectrum shows Balmer and He I emission, which makes this a good CV candidate. There is also consistently strong He II emission with $\frac{{\rm He II 4686}}{{\rm H}\beta} = 2$ in the first epoch of observations and $\frac{{\rm He II 4686}}{{\rm H}\beta} = 0.8$ in the second, suggesting an IP. X-ray observations described in \citet{jon11} show that the X-ray spectrum of CX28 is hard, $\frac{[2.5-8\,keV]-[0.3-2.5\,keV]}{[0.3-8\,keV]}=0.5$, which further supports an IP classification, though there aren't enough photons in the 2 ks observation for a full spectral fit or for timing analysis. Unabsorbed $\frac{F_{X}}{F_{r'}}\sim 0.6$ which is consistent with IPs and CVs. This source cannot be a qLMXB based on the He II emission, and the X-ray strength is too weak for an active LMXB or accretion disk corona. The strength of the He II emission and hardness of the X-ray spectrum suggest that CX18 is an IP, though there are no observations of the defining IP characteristic of an X-ray spin period.  The source varies by 0.7 magnitudes, but is not demonstrably periodic in Mosaic-II data. Observations from June 2011 at the Swope telescope, shown in Fig. \ref{fig:lcfollowup} - panel (d), do not exhibit any period on timescales of minutes to an hour. There is some suggestion of a 2.76 hr period, but the observations do not extend long enough to bear this out as we do not cover multiple periods. If this is the real period, flickering on top of the periodic changes could swamp the signal in our Mosaic-II observations. It is worth noting that this falls in the period gap for CVs and could be further evidence for an IP interpretation because the CV period gap is not observed in IPs.  The changes in RV observed between the VIMOS and NTT observations, shown in Table \ref{tab:lines}, are consistent with velocity variations in either the emitting regions or the motion of the disk around the mass of the system.

\subsection{CXOGBS J173729.1-292804 (CX37)}

The optical spectrum of CX37 shows Balmer, He I, and HeII emission, meaning that the source is not a qLMXB. The relative strength of He I to Balmer lines is consistent with a WD primary, while the strong HeII emission ($\frac{{\rm He II 4686}}{{\rm H\beta}} = 1.1$) rules out a qLMXB (see Table \ref{tab:sort}). Absorbed $\frac{F_{X}}{F_{r'}} \sim 5$, while adjusting for extinction lowers this value to a minimum of $\frac{1}{16}$ at the bulge, which is consistent with CVs and IPs.  There are 37 photons detected in the 2\,ks Chandra observation of this source \citep{jon11}, which is enough to reveal that the X-ray spectrum of CX37 is very hard with $\frac{[2.5-8\,keV]-[0.3-2.5\,keV]}{[0.3-8\,keV]}=0.9$. 
In the Blanco data, CX37 shows non-periodic variability of amplitude 0.15 
magnitudes from the mean and evidence of an eclipse at least 0.9 magnitudes 
deep, as shown in Figure \ref{fig:lc} - panel (d). The other variations are largely due to flickering, as no significant 
periodic modulations are recovered from a periodogram. The lowest data 
point in the Mosaic-II lightcurve is 4.39 times the standard deviation of 
all the 34 observations, suggesting that it is indeed an eclipse rather than a random 
fluctuation from flickering. There is a second data point that is 2.64 times the standard deviation below the mean, which is likely a second eclipse. The eclipsing points are 2.7 
days apart. Also, catching 2 eclipsing points out of 34 
suggests that the eclipse lasts for $\sim 6^{+4}_{-3}\%$ of the period, which is not unusual for CVs \citep{Sulkanen81}. Since the 
nearest observation to the eclipse is separated by only 1.25 hours, that 
suggests a maximum period of $\sim 21$ hours, assuming the two points are from 
eclipses at the same phase. In 4.5 hours of observations with the Swope telescope at LCO (not shown), 
CX37 varies with no periodicity and with no visible eclipse. The LCO observations provide 
a lower bound to the orbital period, implying that the period is some fraction 
$\frac{2.7 {\rm days}}{n}$, $3 \le n \le 14$. 

Each of the emission lines also exhibit large radial velocities, shown in Table \ref{tab:lines}, from $400-600 {\rm km\,s^{-1}}$. It is possible to produce offsets this high in low inclination and short period magnetic systems. 

The hardness of the X-ray spectrum together with the strong He II emission suggests an IP classification for CX37, but as with CX28, X-ray observations capable of detecting a spin period are necessary to make this classification definitively.

\subsection{CXOGBS J174607.6-261547 (CX561)}

The approximate $r'$ magnitude of CX561 is 20.2. The only feature in this spectrum is $H\alpha$. As shown in Fig. \ref{fig:CX561} and Table \ref{tab:lines}, the H$\alpha$ emission line is split with peaks of unequal heights, suggesting significant contribution from a hotspot on the disk. The separation between the peaks is 1420\ km\,s$^{-1}$, which is not high enough to rule out a CV interpretation \citep{Szkody02,Szkody05,Warner03}. The center of the line profile is at 6576.5 \AA, which suggests a velocity along our line of sight of 630 km\, s$^{-1}$. It is possible to produce these speeds for low inclination and short period magnetic or nova-like WD systems if observed at the right phase. Natal kicks routinely produce LMXBs with systemic velocities on this order.  There is no evidence of HeI or HeII lines in the spectrum, but the signal to noise ratio is so low that a typical amount of HeI for a CV could be present and buried in the noise. 

We have only 5 counts in the Chandra data for this source, so we cannot know the shape of the spectrum, but using assumptions in Jonker et al (2011) we find that $\frac{F_{X}}{F_{r'}}$ lies in the range of $\frac{1}{50} \,\mbox{-} 1$, depending on how much of the dust in the line of sight it lies behind. This is consistent with both CVs and qLMXBs. It is too X-ray bright compared to its optical emission to be a coronally active star, and $H\alpha$ is not redshifted enough to suggest an AGN as X-ray faint as CX561 is.  Its absolute magnitude at a distance of 1 kpc would be consistent with quiescent CVs and qLMXBs depending on the donor star, and its absolute magnitude at a distance of 8 kpc is $M_{r'} \approx 1$ which is consistent with the brightest qLMXBs. At bulge distance, $L_X = 6 \times 10^{32}\, {\rm ergs\,s^{-1}}$ which is consistent with both NS and BH qLMXBs, while $L_X = 5 \times 10^{30}\, {\rm ergs\,s^{-1}}$ for a distance of 1 kpc, which is consistent with BH qLMXBs and CVs. If the system is distant, the X-ray to optical flux ratio is a bit low for a NS primary, but this cannot be ruled out for 2 reasons. First, the uncertainty in the ratio is quite high. With only 5 photons in the X-ray and no spectral information, and without photometric standards in the optical or an exact extinction measurement, this value could be off by an order of magnitude when uncertainties from all of these factors are considered.  Also, if this is indeed a qLMXB with a NS primary, it has not undergone recent outbursts and could display different properties in quiescence than those systems that have been followed into quiescence from an active state.

CX561 showed no significant optical variability in the Mosaic-II data.  There is, however, an eclipsing binary with $r'$ magnitude $17.2 \pm 0.5$ with a 12.5 day period next to the emission line object. It is marked in the finder chart in Fig. \ref{fig:finders} with a white circle. It is this nearby variable that has been mistakenly identified as the counterpart by \citet{Udalski12}.

\section{Conclusion}

We have proposed identifications to the optical counterparts to 4 of the 5 initial GBS sources that show emission lines. Of these, 1 is a definite IP, one is a DN, and 2 others are good candidates for IPs. CX5 shows periodic X-ray behavior on the timescale of minutes, strong He II emission, a hard X-ray spectrum, a possible optical orbital period of $\sim 10 \, P_{spin}$, and an X-ray to optical flux ratio just less than unity, making an IP classification certain.  CX18 shows He I and Balmer emission and 2 DN outbursts. CX28 and CX37 both show strong HeII emission with hard X-ray spectra but are too faint in the X-ray to be XRBs, making them likely IPs. The 5th source, CX561, has an uncertain classification without further follow up, though the double peaked H$\alpha$ line means it is a close binary that could either contain a WD or a BH as the compact object. It is not surprising that our large field Mosaic-II observations did not isolate orbital side band periods in the IPs which can be on the order of hundreds to thousands of seconds, but they were sufficient to determine the optical counterpart of identified X-ray sources and to identify eclipsing systems. 

These results represent too small and biased a sample of the survey data to allow any conclusions to be drawn about the population of sources we are finding, but they are consistent with expectations in Jonker et al (2011). We will continue to identify new objects as counterparts are identified and classified.  

\begin{acknowledgements}
This work was supported by the National Science Foundation under Grant No. AST-0908789, by the Louisiana Board of Regents Fellowship, by the NAS/Louisiana Board of Regents grant NNX07AT62A/LEQSF(2007-2010) Phase 3-02. P.G.J. and M.A.P.T. acknowledge support from the Netherlands Organisation for Scientific Research, and the work of their student Oliwia Madj. This research has made use of NASA's Astrophysics Data System Bibliographic Services and of SAOImage DS9, developed by Smithsonian Astrophysical Observatory.
\end{acknowledgements}

{\it Facilities:} \facility{CTIO:1.3m}, \facility{Blanco}, \facility{CXO}, \facility{Swope}, \facility{NTT}, \facility{VLT:Melipal}

\begin{table*}
\begin{center}
\caption{}
\begin{tabular}{l l l c c c c}
\hline
\hline
\noalign{\smallskip}
CX ID  & Line & epoch & $\lambda_{Obs}$ & Radial Velocity & EW & FWHM \\
 & & & ( \AA ) & ${\rm km\, s^{-1}}$ & ( \AA ) & ( \AA )  \\
\noalign{\smallskip}
\hline
5  &  HeII 4686 & 1 & 4682.7 & $-190\pm 70$& $-30 \pm 4$ & 23  \\
 &  H$\beta$ & 1 & 4863.9 &$160\pm 10$ & $-13 \pm 1$ & 12  \\
 &  HeI 5876 & 1 & 5879.3 &$190\pm 20$ & $-4.9 \pm 0.3$ & <14\,\footnotemark[2] \\
 & H$\alpha$ & 1 & 6565.3 & $112\pm 5$ & $-50 \pm 2$ & 13 \\
 &  HeI 6678 & 1 & 6681.1 & $130\pm 20$  & $-4.9 \pm 0.2$ & 9  \\ 
\noalign{\smallskip}
18 &  H$\delta$ & 1 & 4096.4 &$-390\pm 90$ & $-18 \pm 3$ & 18 \\
 &  H$\gamma$ & 1 & 4338.8 &$-110\pm 80$ & $-25 \pm 2$ & 19 \\
 &  HeI 4471 & 1 & 4470.3 &$-80\pm 20$ & $-9 \pm 2$ & <11\,\footnotemark[2] \\
 &  H$\beta$ & 1 & 4858.8 & $-160\pm 50$& $-25 \pm 1$ & 21  \\
 &  HeI 5876 & 1 & 5874.4 &$-60\pm 150$ & $-3 \pm 1$ & 19 \\
 & H$\alpha$ & 1 & 6562.1 &$-30\pm 10$ & $-25 \pm 1$ & 21  \\
 &  HeI 6678 & 1 & 6680.2 & $90\pm 30$& $-2.6 \pm 0.4$ & <11\,\footnotemark[2] \\
\noalign{\smallskip}
28  &  H$\delta$ & 1 & 4101.5 & $-20\pm 10$& $-15 \pm 2$ & 9 \\%
 &  H$\gamma$ & 1 & 4340.4 &$-4\pm 10$ & $-24 \pm 2$ & 19 \\%
 &  Bowen Blend & 1 & 4635.6 &$$ & $-12 \pm$ 1 & 34 \\%
\noalign{\smallskip}
 &  \multirow{2}{*}{HeII 4686} & 1 & 4684.9 &$-50\pm 20$ & $-22 \pm 2$ & 11 \\ 
 &   & 2 & 4691.8 & $390\pm 30$ & $-20 \pm 2$ \\
\noalign{\smallskip}
 &  \multirow{2}{*}{H$\beta$} & 1 & 4858.9 & $-150\pm 20$& $-11 \pm 1$ & 9  \\%
 &   & 2 & 4866.1 & $290\pm 30$& $-24 \pm 5$ & 12 \\%
\noalign{\smallskip}
 &  \multirow{2}{*}{HeII 5412} & 1 & 5416.1 &$250\pm 20$ & $-6.9 \pm 0.6$ 1 & 17 \\%
 &   & 2 & 5418.1 &$360\pm 30$ & $-4.9 \pm 0.3$ & 21  \\
\noalign{\smallskip}
 &  \multirow{2}{*}{HeI 5876} & 1 & 5874.6 & $-50\pm 20$& $-1.4 \pm$ 0.6 & <11\,\footnotemark[2] \\%
 &   & 2 & 5878.1 &$130\pm 20$ & $-3.5 \pm$ 0.3 & 11 \\%
\noalign{\smallskip}
 & \multirow{2}{*}{H$\alpha$} & 1 & 6563.2 &$20\pm 10$ & $-13 \pm 0.5$ & 17 \\%
 &   & 2 & 6565.9 &$140\pm 10$ & $-26.2 \pm 0.3$ & 16 \\%
\noalign{\smallskip}
 &  \multirow{2}{*}{HeI 6678} & 1 & 6683.3 &$230\pm 80$ & $-1.9 \pm 0.6$ & 7 \\
 &  & 2 & 6682.5 &$200\pm 20$ & $-2.6 \pm 0.2$ & 7 \\%
\noalign{\smallskip}
37   &  H$\gamma$ & 1 & 4347.0 &$450\pm 20$ & $-7.5 \pm$ 0.7 & 11 \\
 &  Bowen Blend & 1 & 4626.9 & $$ & $-5.5 \pm$ 0.2 & 28 \\
 &  HeII 4686 & 1 & 4694.9 & $590\pm 30$ & $-16 \pm$ 2 & 14 \\
 &  H$\beta$ & 1 & 4871.1 & $600\pm 10$& $-14 \pm$ 1 & 9 \\
 & H$\alpha$ & 1 & 6574.4 & $530\pm 10$& $-45 \pm$ 2 & 12 \\
\noalign{\smallskip}
561 &  \multirow{2}{*}{H$\alpha$} & 1 & 6553.4, 6585.5\,\footnotemark[1] &$1030\pm 30,-430\pm 30$\,\footnotemark[1] & $-18 \pm 2$, $-31 \pm 2$\,\footnotemark[1] &  \\
 &   &   1 &  6576.6 & $630\pm 40$& $-52 \pm 5$ & 46\\
\hline
\end{tabular}
\caption{Three of the five sources have observed spectra at multiple epochs. Line widths are reported for each epoch in which they are present. Two values separated by commas denote a double peaked line; reported values are for each peak. The FWHM of the lines have been deconvolved with the resolution of the instrument, which is 17\,\AA\ for NTT spectra and 10\,\AA\ for the VIMOS spectrum listed as epoch 2 for CX28. Coordinates of counterparts are from USNO-B1 astrometry.}
\footnotetext[1]{Line is double peaked, center and EW given for each peak. Line is fit with a single Gaussian below.}
\footnotetext[2]{Line widths are upper limits}
\label{tab:lines}
\end{center}
\end{table*}

\begin{table*}
\begin{center}
\caption{}
\begin{tabular}{l c c c c c}
\hline
\hline
\noalign{\smallskip}
CX ID & 5 & 18 & 28 & 37 & 561 \\
\hline
\noalign{\smallskip}
RA (J2000)& 17 40 09.13 &  17 39 35.76 &  17 39 47.01 & 17 37 29.18 & 17 46 07.68 \\
\noalign{\smallskip}
 DEC (J2000) &  $-28\ 47\ 25.7$ & $-27\ 29\ 35.7$ & $-27\ 18\ 08.7$ &  $-29\ 28\ 03.9$ & $-26\ 15\ 49.1$ \\
\noalign{\smallskip}
X-ray Flux  & 157 & 68 &  46 & 37 &  5 \\
(Counts per 2 ks) & & & &  & \\
\noalign{\smallskip}
 Absorbed $\rm{F_{x}}$  &  $2\times 10^{-12}$ & $4\times 10^{-13}$ & $4\times 10^{-13}$ & $3\times 10^{-13}$ & $4\times 10^{-14}$ \\
(ergs\,cm$^{-2}$\,s$^{-1}$) & & & & & \\
\noalign{\smallskip}
 Unabsorbed $\rm{F_{x}}$ &  $4\times 10^{-12}$ & $9\times 10^{-13}$ & $6\times 10^{-13}$ & $5\times 10^{-13}$ &  $7\times 10^{-14}$ \\
(ergs\,cm$^{-2}$\,s$^{-1}$) & & & & & \\
\noalign{\smallskip}
 Absorbed $\rm{F_{r'}}$  &  $2\times 10^{-12}$ & $2\times 10^{-13}$ & $6\times 10^{-13}$ &  $6\times 10^{-14}$ &  $3\times 10^{-14}$ \\
(ergs\,cm$^{-2}$\,s$^{-1}$) & & & & & \\
\noalign{\smallskip}
Unabsorbed $\rm{F_{r'}}$ &  $1\times 10^{-11}$ & $8\times 10^{-12}$ & $3\times 10^{-11}$ &  $8\times 10^{-12}$ &  $3\times 10^{-12}$ \\
(ergs\,cm$^{-2}$\,s$^{-1}$) & & & & & \\
\noalign{\smallskip}
Unabsorbed $\log (\frac{\rm{F_{x}}}{\rm{F_{r'}}})$ &  $-0.4$\,\footnotemark[1] & $0.3$ & $-0.2$ & $0.7$ & $0.1$  \\
\noalign{\smallskip}
$\log (\frac{\rm{F_{x}}}{\rm{F_{r'}}})$ at Bulge & - & $-0.9$ & $-1.7$ & $-1.2$ & $-1.6$ \\
\noalign{\smallskip}
$E(B-V)$ &  1.7 &   1.6 & 1.6 & 2.0 & 1.8 \\
at bulge & & & & & \\
\noalign{\smallskip}
$ \frac{HeII 4686}{H\beta}$ &  2.3 &  $<0.1$ &  2.0,0.83 & 1.1  &  --- \\
\noalign{\smallskip}
$\frac{HeI 4471}{H\beta}$ &  --- &  0.4 &  --- & --- & --- \\
\noalign{\smallskip}
$\frac{HeI 6678}{H\beta}$ &  0.4 &  0.1 &  0.1 &  0.3 & --- \\
\noalign{\smallskip}
References & 1,2,3,5 &  4,5 &  4,5 & 3,5 & 3,5 \\
\hline
\end{tabular}
\tablerefs{ (1) \citet{Sakano00}; (2) \citet{Kaur10}; (3) \citet{Gonzalez11}; (4) Gonzalez et al (2012); (5) Jonker et al (2011)}
\footnotetext[1]{The value for $N_H$ determined by \citet{Kaur10} is used to calculate $E(B-V)$ and flux values for CX5. This likely somewhat overestimates reddening as some of the $N_H$ measured by \citet{Kaur10} is intrinsic to the system.}
\label{tab:fx}
\end{center}
\end{table*}


\begin{thebibliography}{}

\bibitem[Alard \& Lupton(1998)]{Alard98} Alard, C. \& Lupton, R.~H.,\ 1998, \apj,503,325

\bibitem[Alard(2000)]{Alard00} 
Alard, C.\ 2000, \aaps, 144, 363

\bibitem[Bohlin et al.(1978)]{Bohlin78} Bohlin, R.~C., Savage, 
B.~D., \& Drake, J.~F.\ 1978, \apj, 224, 132 

\bibitem[Cardelli, Clayton, \& Mathis(1989)]{Cardelli89} Cardelli, J.~A., 
Clayton, G.~C., \& Mathis, J.~S.\ 1989, \apj, 345, 245 

\bibitem[Casares et al.(1991)]{Casares91} Casares, J., Charles, 
P.~A., Jones, D.~H.~P., Rutten, R.~G.~M., 
\& Callanan, P.~J.\ 1991, \mnras, 250, 712 


\bibitem[Echevarria(1988)]{Echevarria88} Echevarria, J.\ 1988, \mnras, 233, 513 

\bibitem[Edmonds et al.(1999)]{Edmonds99} Edmonds, P.~D., 
Grindlay, J.~E., Cool, A., et al.\ 1999, \apj, 516, 250 

\bibitem[Fender et al.(2003)]{Fender03} Fender, R.~P., Gallo, 
E., \& Jonker, P.~G.\ 2003, \mnras, 343, L99 


\bibitem[Garcia et al.(2001)]{Garcia01} Garcia, M.~R., 
McClintock, J.~E., Narayan, R., et al.\ 2001, \apjl, 553, L47 


\bibitem[Gonzalez et al.(2011)]{Gonzalez11} Gonzalez, O.~A., Rejkuba, M., Zoccali, M., Valenti, E., \& Minniti, D.\ 2011, \aap, 534, A3 

\bibitem[Gonzalez et al.(2012)]{Gonzalez12} Gonzalez, O.~A., Rejkuba, M., Zoccali, M., et al.\ 2012, \aap, 543, A13 


\bibitem[Grindlay(1999)]{Grindlay99} Grindlay, J.~E.\ 1999, 
Annapolis Workshop on Magnetic Cataclysmic Variables, 157, 377 

\bibitem[Grindlay(2006)]{Grindlay06} Grindlay, J.~E.\ 2006, 
Advances in Space Research, 38, 2923 


\bibitem[Hameury et al.(2003)]{Hameury03} Hameury, J.~M., Barret, D., Lasota, J.-P., et al.\ 2003, \aap, 399, 631 

\bibitem[Hynes, Robinson, \& Jeffrey(2004)]{Hynes04} Hynes, R.~I., Robinson, E.~L., Jeffrey, E.\ 2004, \apj, 608,101


\bibitem[Ivanova et al.(2005)]{Ivanova05} 
Ivanova, N., Belczynski, K., Fregeau, J.~M., \& Rasio, F.~A.\ 2005, \mnras, 358, 572 

\bibitem[Izzo et al.(2004)]{Izzo04} Izzo, C., Kornweibel, N., 
McKay, D., et al.\ 2004, The Messenger, 117, 33 


\bibitem[NOAO Data Handbook(2009)]{mosaichandbook}
Jacoby, G., Schweiker, H., Jannuzi, B., et al.\ 2009, NOAO Data Handbook (v.1.1), Ch.2.

\bibitem[Jonker et al.(2011)]{jon11}
 Jonker, P.~G., Bassa, C.~G., Nelemans, G., et al\ 2011 \apjs, 194, 18


\bibitem[Kalogera (1999)]{Kalogera99} 
Kalogera, V.\ 1999, \apj, 521, 723 

\bibitem[Kaur et al (2010)]{Kaur10} Kaur, R., Wijnands, R., 
Paul, B., Patruno, A., \& Degenaar, N.\ 2010, \mnras, 402, 2388 

\bibitem[Lasota (2001)]{Lasota01} Lasota, J.-P.\ 2001, \nar, 45, 449 


\bibitem[Le F{\`e}vre et al.(2003)]{Lefevre03} Le F{\`e}vre, O., 
Saisse, M., Mancini, D., et al.\ 2003, \procspie, 4841, 1670 


\bibitem[Marsh et al.(1994)]{Marsh94} Marsh, T.~R., Robinson, 
E.~L., \& Wood, J.~H.\ 1994, \mnras, 266, 137 


\bibitem[Mauerhan et al.(2009)]{Mauerhan09}
Mauerhan, J.~C., Muno, M.~P., Morris, M.~R., et al.\ 2009, \apj, 703, 30

\bibitem[Menou et al.(1999)]{Menou99} Menou, K., Esin, A.~A., 
Narayan, R., et al.\ 1999, \apj, 520, 276 


\bibitem[Muno et al.(2003)]{Muno03}
Muno, M.~P., Baganoff, F.~K., Bautz, M.~W., et al.\ 2003, \apj, 599, 465


\bibitem[Narayan \& McClintock(2008)]{Narayan08} Narayan, R. \& McClintock, J.~E.\ 2008, NewAR, 51, 733 

\bibitem[Narayan et al.(1997)]{Narayan97}
Narayan, R., Garcia, M.~R., \& McClintock, J.~E.\ 1997, \apj,478, L79

\bibitem[Patterson(1994)]{Patterson94}Patterson, J.\ 1994, \pasp, 106, 209


\bibitem[Pfahl et al.(2003)]{Pfahl03} Pfahl, E., Rappaport, S., 
\& Podsiadlowski, P.\ 2003, \apj, 597, 1036 


\bibitem[Predehl \& Schmitt(1995)]{Predehl95} Predehl, P., \& Schmitt, J.~H.~M.~M.\ 1995, \aap, 293, 889 

\bibitem[Remillard \& McClintock (2006)]{Remillard06} Remillard, R.~A. \& McClintock, J.~E.\ 2006, \araa, 44, 49

\bibitem[Sakano et al.(2000)]{Sakano00} Sakano, M., Torii, K., 
Koyama, K., Maeda, Y., \& Yamauchi, S.\ 2000, \pasj, 52, 1141 

\bibitem[Scaringi et al.(2010)]{Scaringi10} Scaringi, S., Bird, 
A.~J., Norton, A.~J., et al.\ 2010, \mnras, 401, 2207 

\bibitem[(Schlegel et al. 1998)]{Schlegel98} Schlegel, D.~J., 
Finkbeiner, D.~P., \& Davis, M.\ 1998, \apj, 500, 525 

\bibitem[(Servillat et al. 2012)]{champlane12} Servillat, M., 
Grindlay, J., van den Berg, M., et al.\ 2012, \apj, 748, 32 


\bibitem[Shara et al.(2005)]{Shara05} Shara, M.~M., Hinkley, 
S., Zurek, D.~R., Knigge, C., \& Dieball, A.\ 2005, \aj, 130, 1829 


\bibitem[Shaw(2009)]{Shaw09}Shaw, R.~A.,ed.\ 2009, NOAO Data Handbook (Version 1.1; Tucson; National Optical Astronomical Observatory)

\bibitem[Silber(1992)]{Silber92} Silber, A.~D.\ 1992, Ph.D.~Thesis,  


\bibitem[Sulkanen et al.(1981)]{Sulkanen81} Sulkanen, M.~E., 
Brasure, L.~W., \& Patterson, J.\ 1981, \apj, 244, 579 

\bibitem[Szkody et al.(2002)]{Szkody02} Szkody, P., 
G{\"a}nsicke, B.~T., Sion, E.~M., \& Howell, S.~B.\ 2002, \apj, 574, 950 


\bibitem[Szkody \& Henden(2005)]{Szkody05} Szkody, P., \& Henden, A.\ 2005, Journal of the American Association of Variable Star Observers (JAAVSO), 34, 11 


\bibitem[Taam \& Sandquist(2000)]{Taam00}
Taam, R.~E., \& Sandquist, E.~L.\ 2000, \araa, 38, 113

\bibitem[Terzan \& Gosset (1991)]{Terzan91} Terzan, A., \& Gosset, E.\ 1991, \aaps,90,451

\bibitem[Torres et al. (2013)] {Torres13} Torres, M.~A.~P. et al. 2013, in prep

\bibitem[Udalski et al.(2012)]{Udalski12} Udalski, A., Kowalczyk, 
K., Soszy{\'n}ski, I., et al.\ 2012, \actaa, 62, 133 

\bibitem[van den Berg et al.(2009)]{vandenberg09} van den Berg, M., 
Hong, J.~S., \& Grindlay, J.~E.\ 2009, \apj, 700, 1702 


\bibitem[Verbunt et al.(1997)]{Verbunt97} Verbunt, F., Bunk, W.~H., Ritter, H., \& Pfeffermann, E.\ 1997, \aap, 327, 602

\bibitem[Warner(2003)]{Warner03}
Warner, B.,\ 2003,  Cataclysmic Variable Stars, by Brian Warner, ~ISBN 052154209X. ~Cambrige, UK: Cambridge University Press, September 2003


\bibitem[Zurita et al.(2003)]{Zurita03} Zurita, C., Casares, J., 
\& Shahbaz, T.\ 2003, \apj, 582, 369 


\end{thebibliography}
\end{document}